# Extracting the oscillatory component and defining a mean amplitude of thermokinetic oscillations in the H/Pd system

E. Lalik

*Jerzy Haber Institute of Catalysis and Surface Chemistry, Polish Academy of Sciences, Niezapominajek 8, Krakow 30-239, Poland, E-mail: nclalik@cyf-kr.edu.pl*

Abstract

The mean value theorem for integrals has been applied in constructing a base curve for non-equilibrium thermokinetic oscillations, $q(t)$, recorded in oscillatory sorptions of $H_2(D_2)$ in Pd. The mean values are calculated for each period of $q(t)$, followed by cubic spline interpolation, forming the new non-oscillatory curve, $h(t)$, to be used as a baseline for the oscillatory component of the original thermokinetic time series. Crucially, areas, under both $q(t)$ and $h(t)$ are strictly identical. Subsequent pointwise subtraction $q(t)$-$h(t)$ yields another oscillatory time series $g(t)$, considered the oscillatory component extracted from the thermokinetic data. The method has been applied to various experimental time series $q(t)$. Using the $g(t)$ curves, a new parameter, the mean amplitude has been defined and used as a descriptor correlating the intensity of thermokinetic oscillations with various experimental conditions. The mean amplitude turns out to be a linear function of the first ionization potential of noble gases, admixed intentionally to $H_2$ before its contacting Pd.

## 1. Introduction.

Thermokinetic oscillations result from nonlinearity in the rate of heat evolution accompanying chemical reactions far from equilibrium. The resultant time series may offer an insight into their nonlinear mechanism, which can be reflected by the oscillatory parameters, such as the frequency, amplitude and waveform. In the Pd/H system, the thermokinetic oscillations may be measured by instruments such as the Microscal gas flow-through microcalorimeter [1-4]. A vital part of this design is a heat sink that ensures that the heat produced in a solid/gas reaction is instantaneously dissipated, so that the measurements take place isothermally [5]. The oscillatory experiments reveal a succession of waveforms embedded in the ensuing calorimetric curves, manifestly evidencing the underlying degree of nonlinearity. Owing to calibration *in situ*, the obtained curves represent the rate of heat evolution as a function of time. According to standard calorimetric procedures, such curves can be immediately integrated to yield a total amount of heat evolved over a reaction period. Crucially, the heat is related to the change of enthalpy which is a function of state determined only by the initial and final conditions of the system, but not by reaction pathway. On the other hand, the embedded oscillatory component forms a non-standard calorimetric component. It can be characterized by its own parameters, the frequency, amplitude and waveform. Provided that the oscillations are self-sustained, as the chemical oscillations usually are [6,7], these parameters are specifically informed by nonlinearity in the reaction mechanism, but on the other hand, are all independent of the initial conditions by definition [6,8-13]. Thus the frequency and amplitude of thermokinetic oscillations may be investigated in relation to certain reaction parameters, potentially revealing the source of oscillatory behavior.

In our previous experiments with thermokinetic oscillations in the Pd/H system, it was repeatedly confirmed that both the frequency and the amplitude are indeed a function of certain experimental conditions, while at the same time the area under the oscillating calorimetric curves remained invariant with respect to initial conditions, within an error [1,2]. In particular, as the initial conditions, we understand the sample mass and the grain sizes of the Pd powder being used in the sorption. As the reaction different pathways, on the other hand, we consider using admixtures of different inert gases (from He to Kr and $N_2$) to the flow of hydrogen prior to its contact with Pd, necessary to make the thermokinetic oscillations observable microcalorimetrically. With this respect, it was possible to establish a rigorously linear relation between the frequency and a combination of atomic parameters of the inert gases used [2]. These findings support the notion of the observed thermokinetic oscillations to be self-sustainable. As for the amplitude, however, finding a similarly rigorous dependence on experimental conditions turned out more difficult, mainly because of the lack of a well-defined baseline against which the amplitude could be determined. Indeed, the thermokinetic time series obtained with gas flow-through microcalorimeter may not oscillate around the zero axis which in the calorimetric measurements corresponds the thermal equilibrium, i.e., to the null heat transfer. We reiterate that the thermokinetic oscillations occur in a chemical process far from equilibrium, powered by the reaction's ongoing thermal effect. Thus the thermokinetic oscillations are recorded atop of an continuing heat evolution and, as a result, the oscillatory component does not oscillate around any obvious baseline (cf. the red lines in Figs. 1 A and B). While the frequency can still be determined on Fourier transforming [1-3], the amplitude needs a proper baseline against which to be defined. One may argue, however, that a baseline for oscillations must exist, and indeed propose, that the thermokinetic oscillations in the Pd/H system take place around an imaginary non-oscillatory line, that represent a heat evolution accompanying a hypothetical experiment with the same reaction, under entirely identical conditions, proceeding in non-oscillatory manner. This imaginary line must satisfy certain properties. We remember, that since the heat evolving in a reaction is a function of state, so the imaginary non-oscillatory reaction pathway still must be accompanied by the same total thermal effect as the oscillatory one. Consequently, two necessary conditions emerge: (1) the imaginary baseline must be continuous, in order to be integrable, and (2) on integration within respectively the same limits, the area under the imaginary baseline must be the same as that under the experimental time series from which it was derived.

Here we report on a method devised for finding a base curve along the non-equilibrium oscillatory time series, such as those in Figs 1 A and B, that would satisfy the conditions (1) and (2), while not being a simple result of a parallel shifting of the x-axis. We thus seek to construct a curve that would model a "flat", averaged rate of heat evolution in oscillatory reaction, in such a way as to be possibly used as a virtual base-line for the embedded thermokinetic oscillations. However, rather than trying to find such a curve experimentally, we presume that an information that is needed to trace such curve should be already contained in the experimentally obtained oscillatory component. In other words, we expect that a natural oscillation must occur around a certain intrinsic reference level which, while not immediately apparent on visual inspection, nevertheless should be detectable from its record. It seems, that in a vast majority of cases this intrinsic reference line is just identical with the zero-axis, or alternatively it may possibly be approximated by a certain arbitrary function, which is a domain of the signal analysis. In this article, however, we attempt to find the intrinsic reference curve basing not only on a mathematical shape of time series, but also on those aspects of the record that reflect properties of the physical system it represents. Moreover, rather than using a specific function for entire lot, we process the thermokinetic oscillations period by period. The construction is based on mean values, in the sense of the mean value theorem for integrals, found individually for each separate period of oscillations and subsequently interpolated to form the required continuous curve. Mathematically, the procedure begins with discretization of the experimental time series, using the inflection points as the time limits for the resultant segments, each covering a period of time roughly corresponding to the oscillation period, followed by applying the integral mean value theorem to each of these segments separately. The so calculated mean values may be then used as a set of fit points to define a new, non-oscillatory curve (cf. Figures 1 A and B, blue line), purportedly the model for the hypothetical "flat" heat evolution. As a matter of validation, both the modeled flat base-line and the experimental oscillatory time series have to yield the same areas on integration (cf. Figures. 1 A and B), since the total heat of reaction has to be conserved, irrespective of whether oscillatory or not, that is, to satisfy the condition (2).

In satisfying the condition (1), a care has been taken to use a number of interpolated points not only large enough to make the obtained mean value curve continuous, but also to make that number equal to the actual number of points in the original time series, in order to facilitate a subsequent operation of pointwise subtraction. The pointwise subtraction of the two lines yields yet another oscillatory curve that represents the oscillatory component extracted from the original thermokinetic time series. It follows from satisfying the condition (2), that the total area under this extracted oscillatory curve must be zero. Crucially, however, it turns out, that the extracted oscillatory component also preserves the basic topology of the original data (cf. Figures 1 D).

The article is arranged as follows. In Section 3 we illustrate the foundation of the method, using the results of two experiments with the oscillatory sorption respectively of hydrogen or deuterium in the Pd powder. The thermokinetic oscillations accompanying the process of sorption of $H_2$ or $D_2$ in metallic Pd were reported previously in refs. [1-4]. We refer to those two experimental time series as Dataset (1). In Section 4 we then went on to demonstrate how the oscillatory component, once extracted, can be applied in analysis of a further group of nine time series, obtained more recently using different inert gases (Ne, Ar, Kr, Xe, $N_2$) as admixture to hydrogen. The range of nine experiments was designed to ensure, that the observed variations of the frequency and amplitude could be attributed solely to changing the inert gas each time added to hydrogen. To check for reproducibility, the measurements were repeated twice (except for Kr) and are therefore represented each by two close points in the correlation graphs. We refer to the results of those nine experiments as Dataset (2). We use the extracted oscillatory components of the Dataset (2) to define unambiguously a mean amplitude as a single descriptor that can be used as a representation for each of the nine time series. We find a strict linear correlation between the mean amplitudes of the extracted oscillatory components and the first ionization potentials of the inert gases respectively used in recording them.

## 2. Experimental.

The coarse grained palladium powder (purity 99.999%, particle size 0.25–2.36 mm), used for the sorption of $H_2$, and the fine grained Pd powder of granularity less than 75 $\mu$m, used for the sorption of $D_2$, have both been supplied by Aldrich Co.. The following gases: nitrogen (99.999%), hydrogen (99.999%) and deuterium (99.9 %) were provided by Linde Gas Poland S. A. Microscal gas flow-through microcalorimeter, model FMC 4110, has been used for experiments. The design and operation of this instrument has been described in detail in ref. [5]. The experimental procedure leading to periodic oscillatory sorption of $H_2(D_2)$ in Pd has been described in detail in ref. [1,2]. The instrument measures the rate of heat evolution accompanying a solid–gas interaction under isothermal conditions. A sample of Pd powder is placed in a minute microcalorimetric cell (7 mm in diameter, ca. 0.15 $cm^3$ in volume) and the measurement is carried out in a flow-through mode. The cell is located centrally within a much larger metal heat sink. The latter ensures a steady removal of the total of evolving heat and prevents its accumulation within the cell. Once the investigated sample has been sealed within the microcalorimetric cell, the flow of carrier gas through the cell is switched on for the sample to be thermally equilibrated, which is signified by a flat baseline (zero heat transfer) being recorded. The outlet of the cell is equipped with an *in situ* calibrator, located axially within the powdered sample and housing a minute electric coil which makes it possible to produce calibrating pulses of strictly controlled energy and duration. A response to such calibration pulse is recorded as the calibration peak and used to calculate a calibration factor, individually for each experiment. As a reaction is running within the cell, a minute difference of temperatures, between the vicinity of the cell and the locations closer to the outer edge of heat sink, can be measured continuously by a system of thermistors, appropriately located within the latter. All the described time series have been recorded using the Microscal gas flow-through microcalorimeter in our laboratory. However, the data included in Dataset (1) were published previously: the D/Pd results in Ref. [4] and the H/Pd in Ref. [3]. The results shown in Figure 11 were published in Ref. [2], using other figures. The data of Dataset (2) were newly obtained. For those measurements, the experimental set-up was extended by adding a secondary sample of palladium, located down-stream the microcalorimetric cell, as well as a piezoelectric pressure sensor, for concurrent recording of pressure fluctuation. The heat evolution in this secondary Pd sample was not measured. A report on the pressure-related research is currently in preparation. We believe, that these extensions do not have any bearing on the results and conclusions reported in the current contribution.

## 3. Foundation of the concept of extracting the oscillatory component.

**3.1. Construction of the mean value curve.** We will apply the mean value theorem for integrals to our experimental thermokinetic time series $q(t)$ following a discretization procedure. The latter consists of dividing of the whole time series into segments approximately corresponding to the periods of oscillations. Having found the mean value for each segment, we used them to construct a new curve, further referred to as the mean value curve.

The mean value theorem states, that for a function $f(x)$ which is continuous and real-valued within an interval $[a,b]$, there exists a value $c \in (a,b)$ such, that the product $f(c)(b$ - a) equals to the area under the $f(x)$ curve within the $(a,b)$ interval (cf. Figure 2) [14]:

$$f(c)(b-a) = \int_a^b f(x)dx \tag{1}$$

Consider the experimental thermokinetic oscillations $q(t)$ represented in Figure 1A or B (the red lines). After having it divided into a range of segments, we now consider each segment of this time series as a continuous function $q(t)$, $q$: $[t_1,t_2] \rightarrow \mathbb{R}$, but in addition we also assume that $q(t)$ is twice differentiable within the interval $[t_1,t_2]$. The latter condition is necessary, since the limits $t_1$ and $t_2$, are each determined as the inflection points in the ascending parts of the oscillatory curve $q(t)$, marked by the red dots in Figure 1 C. Their positions correspond of the points for which the second time derivative is zero, $d^2q/dt^2 = 0$ (cf. cyan line in Figure 1 C) and the first time derivative $dq/dt$ has a maximum (not shown). With the limits $t_1$ and $t_2$ so defined, each segment approximately corresponds to a single period of oscillations. Using those points as the integral limits, a function mean value $M = q(t_M)$ can be determined for each individual segment:

$$M = \frac{1}{t_2-t_1}\int_{t_1}^{t_2} q(t)dt \tag{2}$$

Apart from the $M$ value, one needs also to determine its abscissa $t_M$, in order to gain a fit point $(t_M, M)$ for the mean value curve. The mean value theorem guarantees that the value $M$ must be somewhere in the segment of the experimental time series between $q(t_1)$ and $q(t_2)$, but it does not determine its abscissa $t_M$, only saying that $t_M \in (t_1,t_2)$. Ideally, one could determine $t_M$ for all segments. However, it would be inefficient to try to find all the $t_M$ values in all segments one by one manually. To simplify the procedure, it would be an attractive option to use a midpoint as the mean value abscissa for each segment. Notably, there are certain functions that have this property that for any interval its mean values abscissa comes at its midpoint, including the linear and certain types of harmonic functions [15]. This is, however, not the case in general. Nevertheless, as a more practical approach, we proceed to approximate the abscissa of $M$ as the midpoint between $t_1$ and $t_2$

$$t^*_i = (t_{2i} - t_{1i})/2 \tag{3}$$

This is illustrated in Figure 1 C. Effectively, using formula (3) recurrently in each ith segment, we define a series of midpoints having coordinates $(t^*_1, M_1)$, $(t^*_2, M_2,)$ ... $(t^*_n, M_n,)$, corresponding to the succession of segments, from 1 to $n$ (cf. blue dots in Figures 1 A and B, the mean values are also listed in Tables 1 and 2). These points are evenly distributed, separated by an extent of time approximately corresponding to the oscillation period (cf. blue dots in Figure 1 D). In general, they lay close, but not exactly on the curve *q*(*t*). Using them as fit points for standard cubic spline interpolation produces a new (non-oscillatory) curve *h*(*t*), running midway across the calorimetric time series (cf. the blue lines in Figures 1A, B and D). As the values *t** are used to approximate the values $t_M$, therefore the curve *h*(*t*) (based on *t** values) is an approximation of the proper mean value curve (that would be based on $t_M$ values). Now we need to test how good this approximation is.

The validity of the approximation of the mean value curve by the curve *h*(*t*) can be confirmed both experimentally and mathematically. The experimental evidence consists of checking whether the areas under the curves *q*(*t*) and *h*(*t*) are equal. Physically, the mean value curve represents a hypothetical time-evolution of heat that the system would produce if the reaction proceeded in a non-oscillatory manner, under the reaction conditions otherwise identical to those of the oscillatory process. As a "sanity check", to assess the closeness of the *h*(*t*) approximation, a condition has to be used, that the total heat evolution represented by the area under this hypothetical non-oscillatory curve must be equal to that under the actual oscillatory time series. Indeed, the integration confirms that these areas are practically equal, found to be 35652 and 35643 mJ, respectively under the curves *q*(*t*) and *h*(*t*) in the sorption of deuterium in Pd (cf. Figure 1 A) and, similarly, 27976 and 27970 mJ in the sorption of hydrogen (cf. Figure 1 B). (cf. also Table 3 for the analogous agreement in Dataset (2)) Thus the experiments confirm the validity of using the curve *h*(*t*) as approximation of the mean value curve. This finding also confirms experimentally the validity of using formula (3), i.e., the approximation of $t_M$ by *t**. We will further analyze the mathematics behind using formula (3) in Section 3.3.

**3.2. Extracting the oscillatory component by the pointwise subtraction of *h*(*t*) from *q*(*t*).** The pointwise subtraction of the mean value curve *h*(*t*) from the thermokinetic oscillations *q*(*t*) yields a new oscillatory curve $g(t) = q(t) - h(t)$. Figure 1 D illustrates this operation for an enlarged fragment comprising three successive segments of the thermokinetic oscillations. The blue dots represents the corresponding three mean value points approximated with formula (3). Figures 1 E and F shows two examples of such pointwise subtraction for the entire experimental time series recorded in the sorption experiments in the $Pd/D_2/N_2$ (Figure 1 E) and the $Pd/H_2/N_2$ system (Figure 1 F). To achieve an effect of juxtaposition, the new *g*(*t*) curve is shown as a fill plot representing oscillations around the zero baseline, against the original time series *q*(*t*). An important requirement to be satisfy, while carrying out the numerical interpolation of the mean values *M*, is that the number of points in the resultant *h*(*t*) curve must be equal to that in the experimental *q*(*t*) time series. It is clear, that the pointwise subtraction preserves the frequency of thermokinetic oscillations. Hence, the *g*(*t*) curve represents the extracted oscillatory component of the original thermokinetic time series *q*(*t*). Expectedly, since the areas under *q*(*t*) and *h*(*t*) are identical, the total area under the curve *g*(*t*) should be zero and this is indeed the case. This aspect will be discussed in more detail in Section 3.4 (cf. also Figure 4).

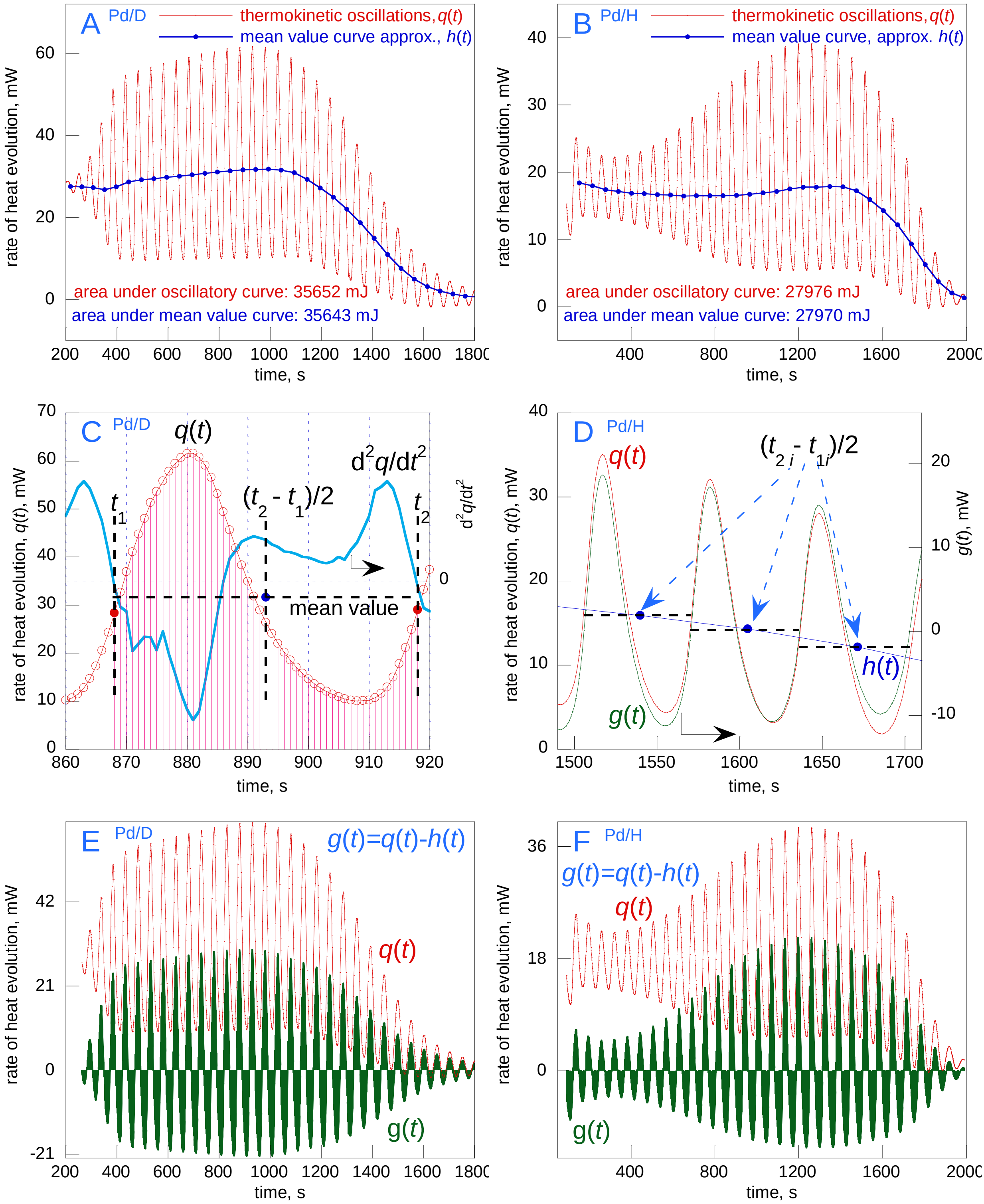

**Figure 1.** **A** and **B)**: The approximated mean value curves $h(t)$ plotted against the thermokinetic oscillations $q(t)$, recorded, respectively, in $Pd/D_2/N_2$ and $Pd/H_2/N_2$ sorption experiments. **C)**: A single segment has its limits at inflection points $t_1$ and $t_2$ (red dots) identified with the zeros of the second derivative $d^2q/dt^2$ curve (cyan). The fit point (blue dot), for the curve $h(t)$ to pass through, is established at the center, $(t_2 - t_1)/2$. **D)**: A succession of three mean value fit points (blue dots), shown against the ensuing oscillatory curve $g(t) = q(t) - h(t)$. Note, that $g(t)$ (green) is plotted against $y2$ axis, shifted with respect to $y$ axis, but of the same scale. **E** and **F)**: The extracted oscillatory curves $g(t)$ (filled green) retain the original frequency of $q(t)$.

**3.3. Mathematical underpinning of using formula (3).** (Remark on notation: To avoid an excessive use of subscripts in the integration limits, for the sake of visual clarity of equations, the letters $a$, $b$ and $c$ are used in this section (also in Figure 2), instead of, respectively, $t_1$, $t_2$ and $t_M$ used elsewhere in this work.)

Mathematically, using formula (3) to approximate the positions of mean values can be justified by the following reasoning. Rewriting equation (1) we obtain

$$\int_a^c f(x)dx - f(c)(c-a) = -\int_c^b f(x)dx + f(c)(b-c). \quad (4)$$

Point $c$ divides the interval (a,b) into two partitions ($a$,$c$) and ($c$,$b$) (cf. Figure 2). Formula (4) shows, that the area between the curve $f(x)$ and the line $y = f(c)$ over the partition ($a$,$c$) and the area between those lines over the partition ($c$,$b$) are equal in absolute value, but opposite in sign (cf. cyan shading in Figure 2). It can be shown, however, that such relation holds for any point that lays on the line $y = f(c)$ over the interval ($a$,$b$), i.e., having an abscissa $d \in (a,b)$, shifted from $c$ by any $|\Delta c|$. Splitting the definite integrals in (4) we obtain:

$$\int_a^d f(x)dx + \int_d^c f(x)dx - f(c)(c-a) = -\int_d^b f(x)dx + \int_d^c f(x)dx + f(c)(b-c) \quad (5)$$

After subsequent addition to both sides of (5) the terms $-f(c)d$ and $f(c)c$ and after cancellation we have

$$\int_a^d f(x)dx - f(c)(d-a) = -\int_d^b f(x)dx + f(c)(b-d) \quad (6)$$

which can be rewritten as

$$\int_a^d f(x)dx - f(c)\int_a^d dx = -\int_d^b f(x)dx + f(c)\int_d^b dx \quad (7)$$

or

$$\int_a^d [f(x) - f(c)]dx = -\int_d^b [f(x) - f(c)]dx. \quad (8)$$

The arbitrary point $d \in (a,b)$ divides the interval ($a$,$b$) into two partitions,($a$,$d$) and ($d$,$b$). The areas enclosed between the curve $f(x)$ and the line $y = f(c)$ within those intervals, respectively, represented by LHS and RHS of (8), are equal in absolute value, but opposite in sign.

The relation (8) holds irrespective of selection of $d$. It follows, that the definition of $t^*$ according to formula (3) does not violate the conditions of the mean value theorem (1), in fact, $t^*$ is a special case of $d = (b - a)/2$. Hence, the mean value $M$ once calculated using formula (2), defines a horizontal line, that, technically speaking, can be used as a lifted $x$-axis from which to integrate the function $q(t)$.

It can further be shown, that both sides of (8) attain an extremum for $x = c$. To see this, note, that since $d$ can be selected anywhere within the interval ($a$,$b$), the expression (8) can be rewritten using the integrals with variable upper limit $u$, instead of the fixed $d$

$$\int_a^u [f(x) - f(c)]dx = -\int_u^b [f(x) - f(c)]dx \quad (9)$$

or

$$\int_a^u [f(x) - f(c)]dx = \int_b^u [f(x) - f(c)]dx \quad (10).$$

Now we differentiate both side with respect to $u$ and equate the result to zero,

$$\frac{d}{du}\int_a^u [f(x) - f(c)]dx = \frac{d}{du}\int_b^u [f(x) - f(c)]dx = 0 \tag{11}$$

yielding $f(x) = f(c)$, which is only true for $x = c$.

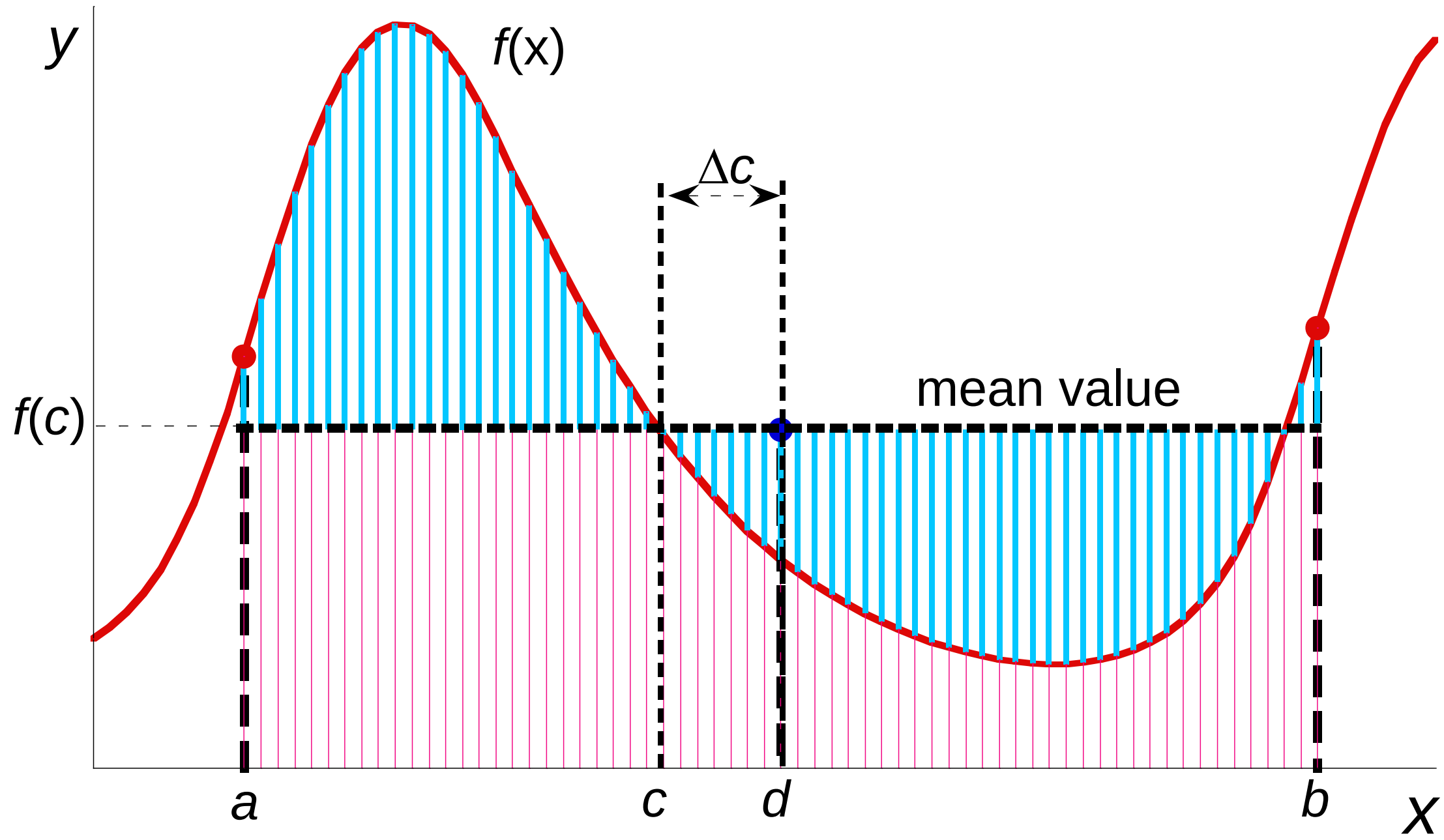

**Figure 2**. The mean value theorem for integrals (equation(1)) makes it possible to find a mean value point $f(c)$ that exists in the segment of the curve $f(x)$ over the interval $(a,b)$, but it does not determines its abscissa $c$. The cyan-shaded partitions above and below the mean value $f(c)$ are equal, irrespective of whether the division point for the segment $(a,b)$ is $c$ or $d$. The relation (8) shows that the same holds for any $x \in (a,b)$ (cf. Section 3.3). For the sole purpose of constructing the mean value curve, the abscissa of the mean value $f(c)$ can be approximated by a midpoint $d = (b - a)/2$.

**3.4. Topological analysis of the pointwise subtraction of *h*(*t*) from *q*(*t*).** The relation (8) reappears in Figure 3 illustrating its implementation to actual experimental data. Focusing on a single segment of $q(t)$, the figure represents topologic details of extracting of the oscillatory component $g(t)$ (in green) from the calorimetric time series $q(t)$ (in red). The single segment in Figure 3 A is delimited by the abscissa values $t_1$ and $t_2$ of the consecutive inflection points in $q(t)$, marked as red dots. The blue dot marks the point of coordinates $(t^*, M)$, with $t^* = (t_2 - t_1)/2$, which is the fit point for the interpolated curve $h(t)$ (in blue) to pass through in this segment. Note, that this fit point is shifted by $\Delta t$ from the actual position $t_M$ of the mean value $M$ (an analogue to $\Delta c$ in Figure 2). The resultant green line $g(t) = q(t) - h(t)$ is plotted against the right hand side axis $y2$, adjusted in such a way that the zero in the $y2$ axis corresponds to $M$ in the $y$ axis, with both axes $y$ and $y2$ having the same scale. Physically, both the axes $y$ and $y2$ show the rate of heat evolution in mW units. Since $q(t)$ represents the rate of heat evolution as a function of time, so the red shading represents the total heat evolved during the period from $t_1$ to $t_2$. The cyan shading illuminates the area between the curve $q(t)$ and the line $y = M$, yielded by integration of the difference $[q(t) - M]$ from $t_1$ to $t_2$. Similarly to the analogous area in Figure 2, the "cyan area" in Figure 3 A is also divided into two partitions, respectively within intervals ($t_1$,$t$*) and ($t$*,$t_2$). Physically, it may be considered as a virtual flow of energy above and below the mean value level. Relation (8) applies and, accordingly, the areas of these (cyan) partitions in Figure 3 A are equal in absolute value, but opposite in sign. By virtue of (8), this must be true for each segment of the time series $q(t)$. So there is a detailed balance within each segment between the virtual energy production above and below the mean value. This is visualized by histograms in Figure 4 A and C. Each double bar represents a segment. The black half represents the integral of $[q(t) – M]$ within interval ($t_1$,$t$*), yielding the energy above the mean value $M$. The cyan half represents the absolute value of this integral within ($t$*,$t_2$), yielding the energy below $M$. The absolute values were used to enable visual comparability. In both histograms (respectively for the sorption of $D_2$ and $H_2$ in Pd), the pairwise black-cyan equivalence is evident for entire experiments. The exact data used to create the histograms in Figure 4 are listed in Tables 1 and 2 in columns (b) and (c). These data are an experimental manifestation of relation (8).

In Figure 3 B, the green shading have been added to accentuate the area under the extracted oscillations curve $g(t)$, plotted against the $y$2 axis. Calorimetrically, it can be viewed as consisting of the exothermic and endothermic peaks, with respect to a now linear baseline, i.e., the $g(t) = 0$ line. A visual inspection suggests that the exo and endo “green” peaks should be topologically related to the positive and negative “cyan” lobes of the $[q(t) – M]$ area (the cyan shaded areas are mostly overlapped here, but fully visible in Figure 3 A). It can be shown that both the “cyan” and the “green” features are indeed approximately equal in their areas. But in order to prove this relation, we need first to identify an additional relation between the area of those green peaks, on one hand, and yet another feature, namely: the area enclosed between the segments of the curves $q(t)$ and $h(t)$ limited by points A and B marked in Figure 3 C.

Figure 3 C shows the groundwork of the pointwise subtraction of $h(t)$ from $q(t)$ demonstrated for a single segment AB. Note, that this operation can be viewed as a transformation of the area enclosed between the AB segments of the curves $q(t)$ and $h(t)$ into the area under the CD segment of the curve $g(t)$ (plotted against the $y$2 axis). Each point in the $q(t)$ curve (red crosses) undergoes a vertical shift (downward or upward) to a new position, forming the new curve $g(t)$ (green open dots). It can be shown, that for any point X in the AB segment of the curve $q(t)$, the absolute value of its shift $|q(t_x) – g(t_x)|$ (upward or downward) to its new position X’ in the CD segment of $g(t)$ is a linear function of its abscissa $t_x$ within the interval $[t_A,t_B]$. The index “$x$” indicates, that the variable $t_x$ is only defined within a single segment. In fact, each point in the area between the graphs $q(t)$ and $h(t)$ undergoes such shift, to form the area under the $g(t)$. Therefore, for any such point, having the abscissa $t_x \in [t_A,t_B]$, the pointwise subtraction can be written as

$$g(t_x) = q(t_x) – h(t_x) \tag{12}$$

Note, that the AB segment of $h(t)$ can be approximated as a straight line. Therefore, since $h(t_x)/(t^* – t_x) = \mathrm{tg}\theta$, and using the approximating $\mathrm{tg}\theta = \theta$ we have

$$g(t_x) = q(t_x) – \theta\,(t^* – t_x) \tag{13}$$

so for any point $t_x$ the difference between the corresponding values of the curves $q(t)$ and $g(t)$

is a linear function of $t_x$:

$$\left| q(t_x) - g(t_x) \right| = \left| -\theta\, t_x + \theta\, t^* \right|. \qquad (14)$$

Hence the operation of the pointwise subtraction may in fact be viewed geometrically as a vertical shear of the area enclosed by the AB sections of the $h(t)$ and the $q(t)$, at a small shear angle $-\theta$. Accordingly, the abscissa is being preserved $t' = t$ and the ordinate transformed linearly $y' = -\theta\, t + y$. In matrix notation we have

$$\begin{bmatrix} t' \\ y' \end{bmatrix} = \begin{bmatrix} 1 & 0 \\ -\theta & 1 \end{bmatrix} \begin{bmatrix} t \\ y \end{bmatrix} \qquad (15)$$

and since the determinant of the transformation matrix is one,

$$\left| det \begin{vmatrix} 1 & 0 \\ -\theta & 1 \end{vmatrix} \right| = 1 \qquad (16)$$

therefore the transformation in Figure 3 C is area preserving. It thus preserves the relations between the concave and convex partitions of the area enclosed by the AB sections of $q(t)$ and the AB section of $h(t)$ within the interval $[t_A, t_B]$. Now, the areas of these concave and convex partitions approximately correspond to the areas of the "cyan" lobes (positive and negative) within the interval $[t_1, t_2]$ in Figure 3 A and B. For small deviation between the limits of these intervals $[t_1, t_2]$ and $[t_A, t_B]$, this correspondence is close enough for relation (8) to hold also for the area enclosed by the AB sections of $q(t)$ and $h(t)$, and hence by virtue of (8) the concave and convex partitions are approximately equal to one another in absolute values. This relation, therefore, must be also preserved between the exo and endo peaks in the curve $g(t)$, as they are yielded by the area preserving vertical shear. Hence the important property of the resultant curve $g(t)$ turns out to be the pairwise equivalence of the subsequent exo- and endo-peaks, which also can be demonstrated experimentally (cf. Figure 4 B and D). In the Figure 4 B the filled plot of the $g(t)$ curve (green) is placed alongside the curve representing the integrals of $g(t)\mathrm{d}t$ with variable upper limits (magenta). The total green area, that is, the sum of all exo and endo peaks in the $g(t)$ curve, is balanced to zero and this is reflected in the magenta line, i.e., the integral curve, reaching zero at the end. Remarkably, however, the magenta line hits the zero every time that a new period starts in $g(t)$, as it is illustrated in the enlarged fragment in Figure 4 D. This is an evidence of detailed exo/endo balance being kept in each individual period of the extracted oscillatory curve $g(t)$. Tables 1 and 2 list the exact values of the integrated exo and endo peaks in the curves $g(t)$ in columns (d) and (e).

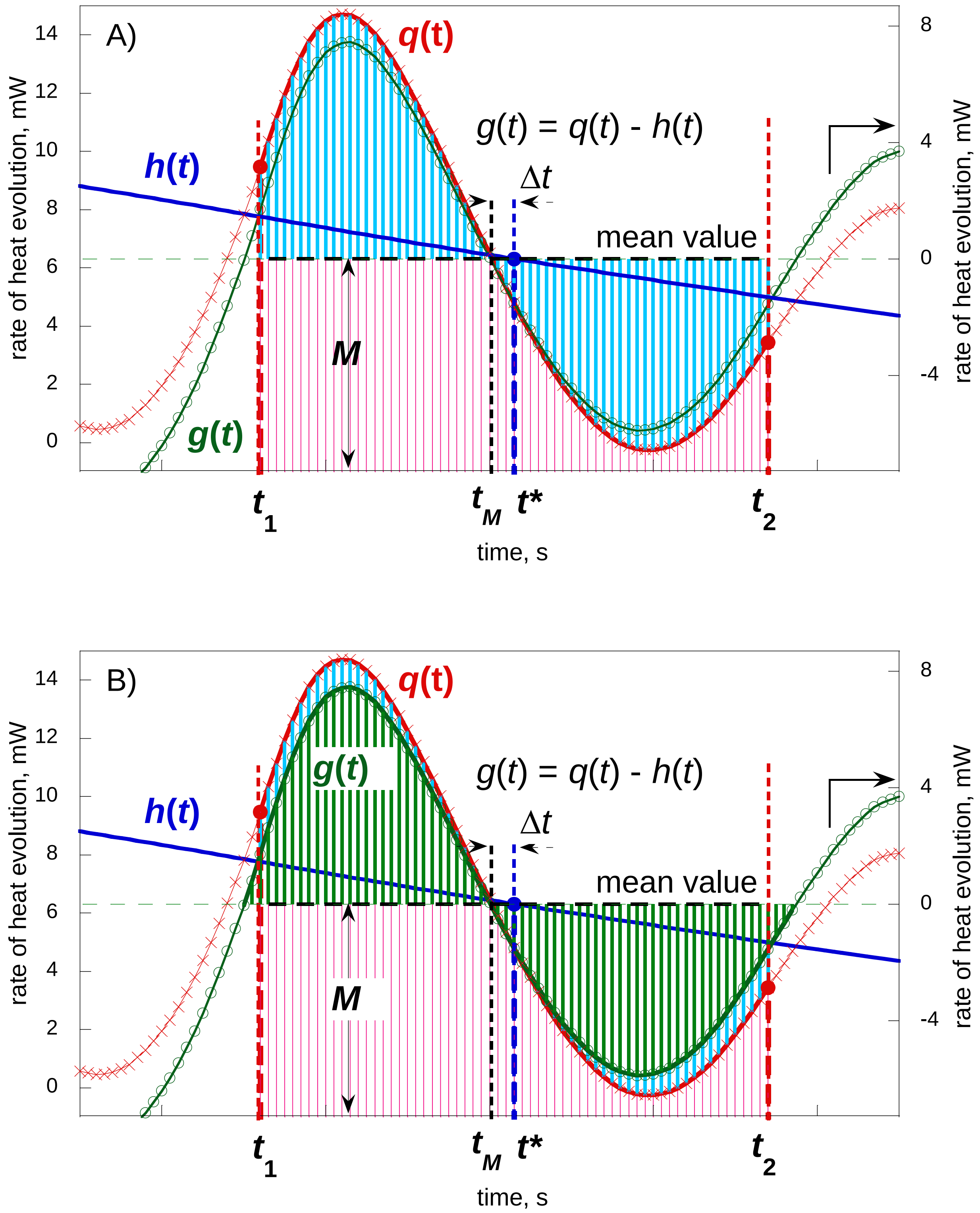
A)
q(t)
g(t) = q(t) - h(t)
h(t)
Δt
mean value
M
g(t)
rate of heat evolution, mW
14
12
10
8
6
4
2
0
8
4
0
-4
t1
tM
t*
t2
time, s
B)
q(t)
g(t)
g(t) = q(t) - h(t)
h(t)
Δt
mean value
M
rate of heat evolution, mW
14
12
10
8
6
4
2
0
8
4
0
-4
t1
tM
t*
t2
time, s

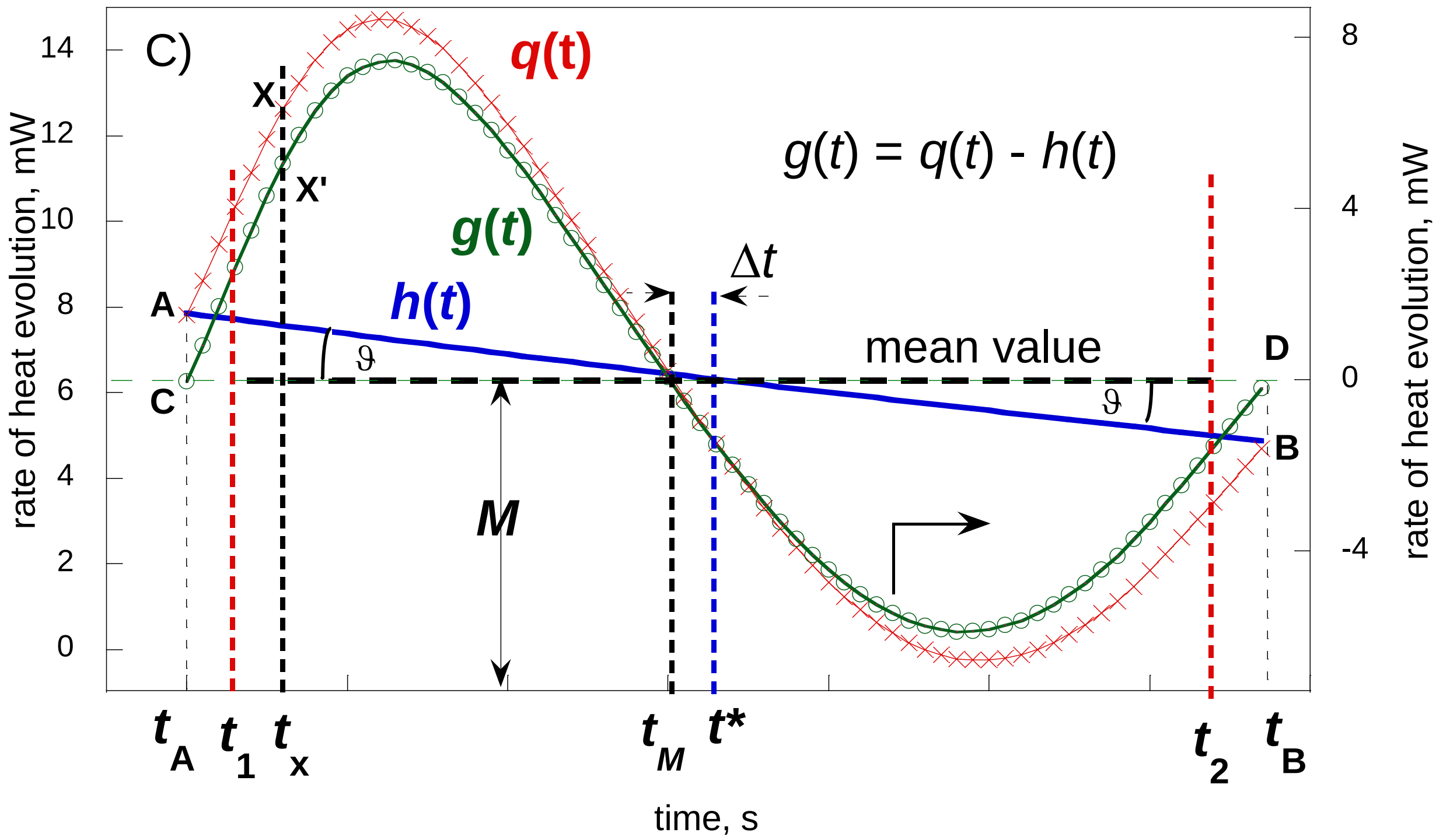

**Figure 3**. Topology of the pointwise subtraction of the approximated mean value curve *h*(*t*) from the thermokinetic curve *q*(*t*) illustrated for a single segment. **A**) The abscissa for the mean value *M* approximated as *t**and used to build the curve *h*(*t*) subtracted from the *q*(*t*). The "cyan" areas between *q*(*t*) and the horizontal line *q* = *M* are equal but of opposite signs. **B**) The affinity of the "cyan" areas to the "green" peaks of the extracted oscillatory curve *g*(*t*). **C**) The *q*(*t*) – *h*(*t*) subtraction may be viewed as a vertical shear of the area enclosed by the AB sections of the *h*(*t*) and the *q*(*t*), at a small shear angle –*θ*, yielding the two peaks of the curve *g*(*t*).

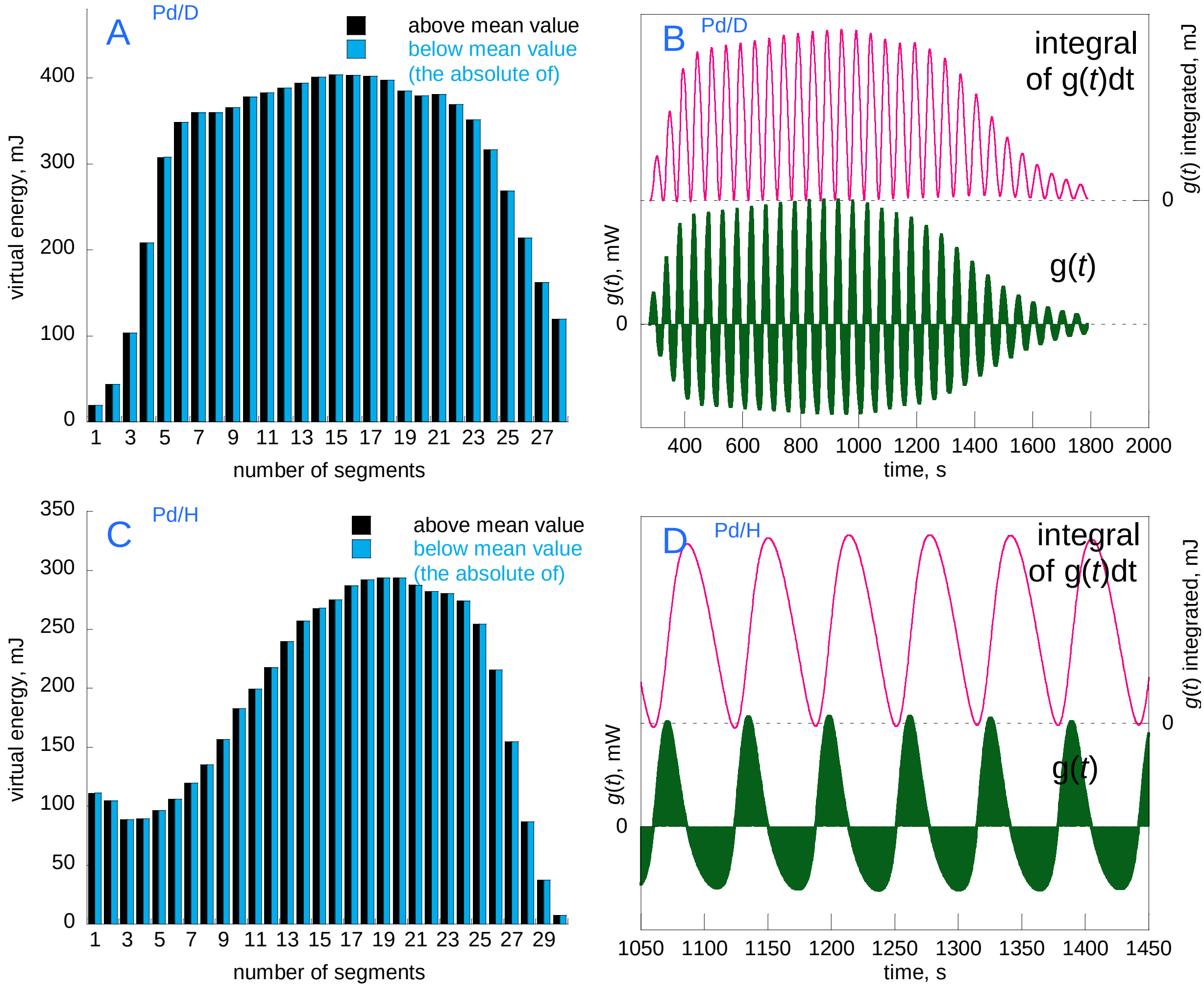

**Figure 4.** The experimental demonstration of formula (8). The histograms in panels **A** and **C** illustrate the equilibrium of the "cyan" partitions of Figure 3 A (i.e., the cyan shaded areas above and below the mean value *M* in Fig. 3 A) for the entire *q*(*t*) curves in Pd/$D_2$ (panel **A**) and Pd/$H_2$ (panel **C**). Each of the black/cyan double bars corresponds to a single segment of *q*(*t*); for swift comparison, the cyan bars are in absolute values. The panels **B** and **D** show the extracted oscillatory curves *g*(*t*) (green) along with their (cumulative) integrals with variable upper limit (magenta), evidencing the detailed exo/endo balance being maintained in each individual periods of the *g*(*t*) curve. Panel **B** shows the entire Pd/$D_2$ sorption; panel **D** shows a 400 s fragment of the Pd/$H_2$ sorption.

Table 1. $Pd/D_2$ results: (a) The mean values obtained for successive segments of the thermokinetic oscillations $q(t)$, used to construct the approximated mean value curve $h(t)$ shown in Figure 1 A (in blue). The values in columns (b) and (c) correspond to the black and cyan bars of histogram in Figure 4 A. The values in columns (d) and (e) represent the areas of the exo and endo peaks of the extracted oscillatory curve $g(t)$ shown (in green) in Figures 1E and 4 B.

| (a)<br>$Pd/D_2$ Mean value, mW | (b)<br>Partitions above mean value, mJ | (c)<br>Partitions below mean value, mJ | (d)<br>Exo peaks in $g(t)$, mJ | (e)<br>Endo peaks in $g(t)$, mJ |
|---|---|---|---|---|
| 27.577 | 19.081 | -19.099 | | |
| 27.508 | 43.517 | -43.496 | | |
| 27.252 | 103.11 | -103.13 | 107.09 | -107.44 |
| 26.780 | 208.22 | -208.21 | 215.89 | -217.54 |
| 27.450 | 307.68 | -307.70 | 320.72 | -320.98 |
| 28.638 | 348.20 | -348.19 | 361.32 | -358.24 |
| 29.176 | 359.50 | -359.50 | 368.39 | -368.20 |
| 29.468 | 359.50 | -359.50 | 373.92 | -373.87 |
| 29.810 | 365.86 | -365.87 | 379.84 | -380.97 |
| 30.074 | 377.98 | -378.00 | 385.92 | -385.59 |
| 30.430 | 382.93 | -382.91 | 392.00 | -391.55 |
| 30.768 | 388.32 | -388.33 | 398.90 | -398.93 |
| 31.106 | 394.16 | -394.15 | 402.85 | -403.84 |
| 31.348 | 400.85 | -400.85 | 408.81 | -408.66 |
| 31.585 | 403.61 | -403.60 | 411.58 | -410.86 |
| 31.693 | 403.02 | -403.03 | 412.66 | -412.16 |
| 31.752 | 402.01 | -402.02 | 410.27 | -410.36 |
| 31.488 | 397.43 | -397.45 | 402.33 | -400.68 |
| 30.884 | 385.05 | -385.03 | 386.86 | -384.11 |
| 29.305 | 379.34 | -379.32 | 375.29 | -373.52 |
| 27.174 | 380.89 | -380.87 | 374.36 | -374.61 |
| 24.928 | 369.10 | -369.10 | 358.95 | -357.18 |
| 22.021 | 351.28 | -351.26 | 334.87 | -334.31 |
| 18.706 | 316.39 | -316.41 | 296.27 | -293.85 |
| 14.875 | 268.55 | -268.56 | 245.18 | -244.76 |
| 10.987 | 213.91 | -213.93 | 190.26 | -191.65 |
| 7.5472 | 162.19 | -162.19 | 142.15 | -144.50 |
| 4.9702 | 119.70 | -119.70 | 105.56 | -107.30 |

Table 2. $Pd/H_2$ results: (a) The mean values obtained for successive segments of the thermokinetic oscillations *q*(*t*), used to construct the approximated mean value curve *h*(*t*) shown in Figure 1 B (in blue). The values in columns (b) and (c) correspond to the black and cyan bars of histogram in Figure 4 C. The values in columns (d) and (e) represent the areas of successive exo and endo peaks of the extracted oscillatory curve *g*(*t*) shown (in green) in Figures 1 F and 4 D (fragment).

| (a) $Pd/H_2$ Mean value, mW | (b) Partitions above mean value, mJ | (c) Partitions below mean value, mJ | (d) Exo peaks in *g*(*t*), mJ | (e) Endo peaks in *g*(*t*), mJ |
|---|---|---|---|---|
| 18.452 | 111.02 | -111.05 | 112.73 | -109.90 |
| 17.999 | 104.59 | -104.59 | 102.71 | -102.45 |
| 17.397 | 88.786 | -88.785 | 87.133 | -89.662 |
| 17.129 | 89.284 | -89.258 | 90.893 | -91.493 |
| 16.899 | 96.520 | -96.549 | 99.087 | -99.157 |
| 16.829 | 106.14 | -106.11 | 109.60 | -110.93 |
| 16.687 | 119.78 | -119.77 | 126.80 | -127.45 |
| 16.618 | 135.04 | -135.03 | 145.35 | -147.52 |
| 16.483 | 156.57 | -156.54 | 169.65 | -167.47 |
| 16.527 | 182.67 | -182.67 | 194.61 | -195.37 |
| 16.550 | 199.62 | -199.59 | 218.07 | -221.94 |
| 16.502 | 217.42 | -217.44 | 244.23 | -244.12 |
| 16.574 | 239.80 | -239.78 | 268.02 | -266.41 |
| 16.754 | 257.10 | -257.08 | 286.10 | -286.97 |
| 16.937 | 267.86 | -267.88 | 300.47 | -301.30 |
| 17.153 | 275.08 | -275.06 | 310.75 | -311.79 |
| 17.534 | 286.96 | -286.97 | 321.92 | -318.38 |
| 17.792 | 291.92 | -291.95 | 323.31 | -324.40 |
| 17.819 | 293.46 | -293.43 | 324.70 | -322.18 |
| 17.912 | 293.61 | -293.64 | 320.99 | -321.35 |
| 17.853 | 287.47 | -287.44 | 314.75 | -314.11 |
| 17.271 | 282.04 | -282.06 | 306.08 | -303.88 |
| 15.929 | 280.39 | -280.39 | 298.02 | -294.92 |
| 14.305 | 274.13 | -274.11 | 284.60 | -283.76 |
| 12.179 | 254.27 | -254.24 | 258.71 | -256.42 |
| 9.3355 | 215.55 | -215.54 | 208.83 | -205.80 |
| 6.3099 | 154.82 | -154.82 | 136.64 | -139.20 |
| 3.7687 | 86.673 | -86.671 | 70.203 | -74.440 |
| 2.0762 | 37.262 | -37.264 | 30.542 | -32.475 |
| 1.3049 | 7.4657 | -7.4675 | | |

## 4. Application of the proposed method to a group of experimental thermokinetic time series (Dataset (2)).

**4.1. General remarks.** As a case in point, we use a database of nine experimental time series (Dataset (2)) representing oscillatory rate of heat evolution recorded with gas flow-through microcalorimeter during the sorption of hydrogen in powdered Pd. A crucial condition for these oscillations to turn observable is an admixture of ca. 10 % vol. of an inert (Ne, Ar, Kr, Xe or $N_2$), added to the stream of $H_2$ prior to its contact with Pd. The range of nine experiments was designed to ensure, that the observed variations of the frequency and amplitude could be attributed solely to changing the inert gas each time added to hydrogen. To check for reproducibility, the measurements were repeated twice (except for Kr).

Figure 5 A illustrates the increasing intensity of thermokinetic oscillations as a function of inert admixture: Ne, Ar or Xe. For clarity, the experiments with Kr and $N_2$ admixtures was not included in the figure. With other reaction conditions strictly preserved, the oscillatory parameters must be related to the inert gas properties. Figure 1 B shows an empirical relation for the frequency which turns out to be a linear function of a product involving the atomic properties, namely, the first ionization potential (FIP) and the square root of the atomic mass (SQRM). A similar relation previously reported in Ref. [2] involved averaged values of frequencies and did not include Xe. Physically, it indicates that the effect that leads to the thermokinetic oscillations likely involves adsorption of the inert gases on the Pd surface. Methodologically, it also confirms a validity of using the frequency as a descriptor unambiguously representing each time series in a functional relation such as the one shown in Figure 5 B. However, Figure 5 A also shows that a possible definition of an analogous descriptor for the amplitude must be more involved, since the amplitude changes in time. In this section, we propose a solution that accounts for this impediment. We begin by extracting the oscillatory components (in a way described in Section 3) from the non-equilibrium oscillations, illustrated in Figure 5 A. Then we go on using the so obtained oscillatory component to define a new parameter, called the mean amplitude and show that it can be used as the desired amplitude-related descriptor, able to characterize unambiguously the original thermokinetic time series.

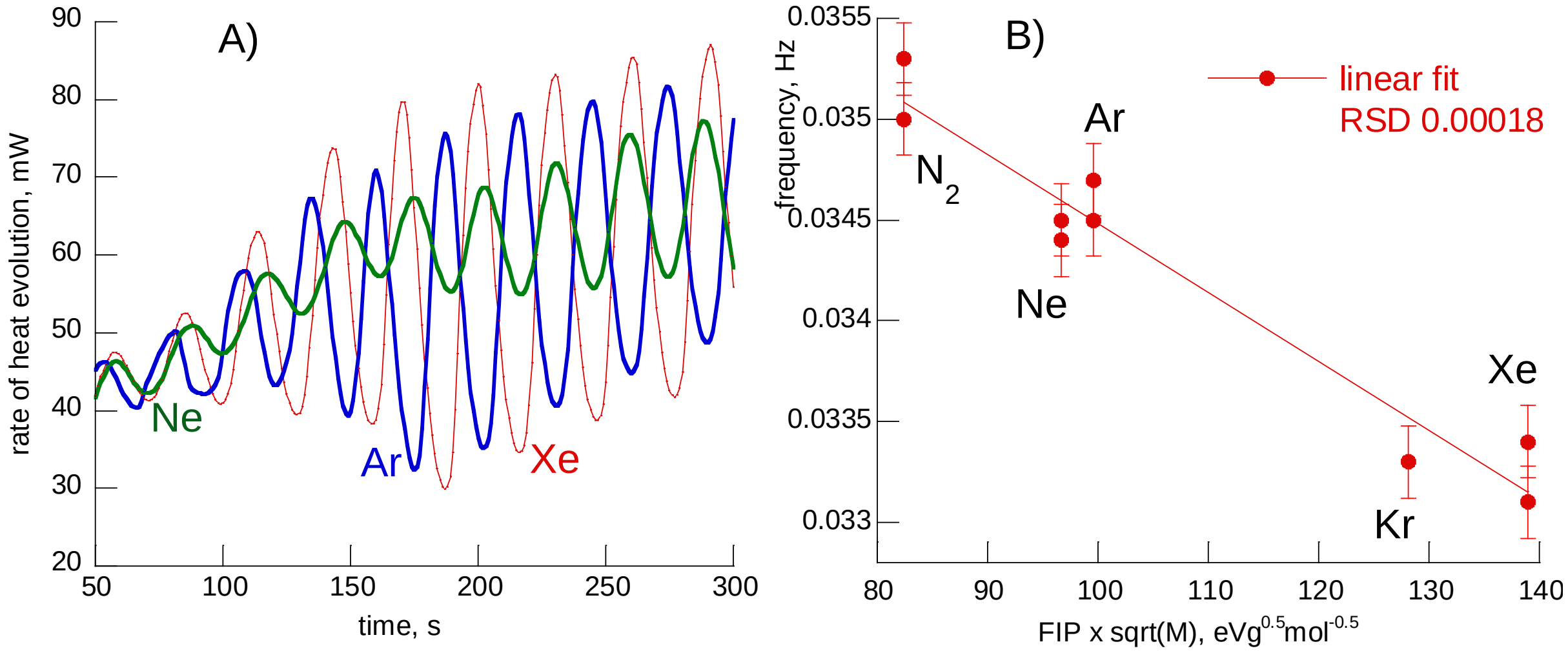

**Figure 5. A)** The increasing intensity of thermokinetic oscillations accompanying the sorption of hydrogen in Pd measured on admixing of Ne, Ar, or Xe inert gas into the $H_2$ stream; for clarity, only three examples of a nine-strong database are shown. **B)** The linear dependence of the oscillation frequency on the product of the first ionization potential (FIP) and the square root of the atomic mass M of the inerts Ne, Ar, Kr, Xe and $N_2$). The bars represent the residual standard deviation (RSD).

### 4.2. Extraction of oscillatory components in Dataset (2).

We note, that, technically, the method of oscillatory component extraction described in Section 3 may be viewed as a two-staged procedure. Stage **(1)** sees the mean value theorem for integrals applied to the experimental oscillatory time series $q(t)$ in order to establish a mean value curve $h(t)$, the latter considered as the imaginary baseline for oscillations. In stage **(2)**, the pointwise subtraction of the two curves, $q(t) - h(t)$ yields a new oscillatory time series $g(t)$ that turns out to be the desired oscillatory component extracted from the experimental time series $q(t)$.

**4.2.1. Stage (1): Constructing the mean value curves and their validation.** To define the mean value curves, first the experimental time series $q(t)$ have to be time-sliced (discretized) into a range of segments, before applying the integral mean value theorem to each one separately. Figure 6 illustrates this procedure for the example of the results obtained for Xe, i.e., the $q_{Xe}(t)$ curve. This involves using the inflection points located in the ascending parts of the oscillatory curve (black diamonds in Figure 6) as the integral limits. Consequently, each segment corresponds to a single period of oscillations. The positions of the resultant mean values for each period, with their abscissa approximated as the middle points for each segments, are shown in Figure 6 as the blue dots. The cubic spline interpolation of those points produces a new curve $h_{Xe}(t)$ running midway across the calorimetric time series (blue curve in Figure 6). Physically, this mean value line $h(t)$ represents a hypothetical curve that the system would trace, if the reaction proceeded in a non-oscillatory manner, under reaction conditions otherwise identical to those of the oscillatory process. Accordingly, the total heat evolution represented by the area under this hypothetical non-oscillatory curve $h(t)$ must be equal to that under the actual oscillatory time series $q(t)$[5]. Indeed, integration of both the curves $q_{Xe}(t)$ and $h_{Xe}(t)$ in Figure 6 yields, respectively, 17816 and 17780 mJ. This is a crucial, physical test for the procedure validity. Table 3 confirms the same good agreement of the $q(t)$ and $h(t)$ areas ($Q$ and $H$) for the whole database of nine experiments presented in this work; the one actually presented in Figure 6 is listed at position Xe(1) in Table 3. The deviation, calculated as $D = \frac{|Q-H|}{0.5(Q+H)}100$, never exceeds 0.9 %.

Table 3. The agreement of the areas under the $q(t)$ and $h(t)$ curves.

| Inert gas admixture (Exp. No) | $Q$ Area under $q(t)$ curve, mJ | $H$ Area under $h(t)$ curve, mJ | Interval of integration, s | $D$ Deviation, % |
|---|---|---|---|---|
| Ne(1) | 18110 | 18097 | 327 | 0.071809 |
| Ne(2) | 15769 | 15676 | 260 | 0.59151 |
| Ar(1) | 18117 | 18102 | 346 | 0.082829 |
| Ar(2) | 17995 | 17979 | 347 | 0.088953 |
| Kr | 17157 | 17307 | 299 | 0.87047 |
| Xe(1) | 17816 | 17780 | 361 | 0.20227 |
| Xe(2) | 17482 | 17369 | 328 | 0.64847 |
| $N_2$(1) | 17650 | 17617 | 363 | 0.18714 |
| $N_2$(2) | 17631 | 17596 | 358 | 0.19871 |

**4.2.2. Stage (2): The pointwise subtraction.** Figure 6 illustrates the extraction of oscillatory component for one of the time series recorded using Xe as the admixture to $H_2$. The pointwise subtraction of $q(t) - h(t)$ yields a new oscillatory time series $g(t)$ that may be considered as an oscillatory component extracted from the experimental thermokinetic time series $q(t)$. Since it oscillates around the zero line, so calorimetrically, it is a succession of alternating exo and endo peaks (cf. Figure 7 B). The mathematical properties of the so obtained $g(t)$ curve has been described in detail in Section 3. It has the same frequency as the original $q(t)$. Since the areas under both $q(t)$ and $h(t)$ are equal, so the total area under $g(t)$, shaded in pink in Figure 6, must be zero. Not only the total area, however, but also within each individual period of the $g(t)$ curve, the consecutive exo and endo peak were proved to be equal in absolute values, but opposite in sign (irrespective of the possible different shapes of the exo and endo peaks).

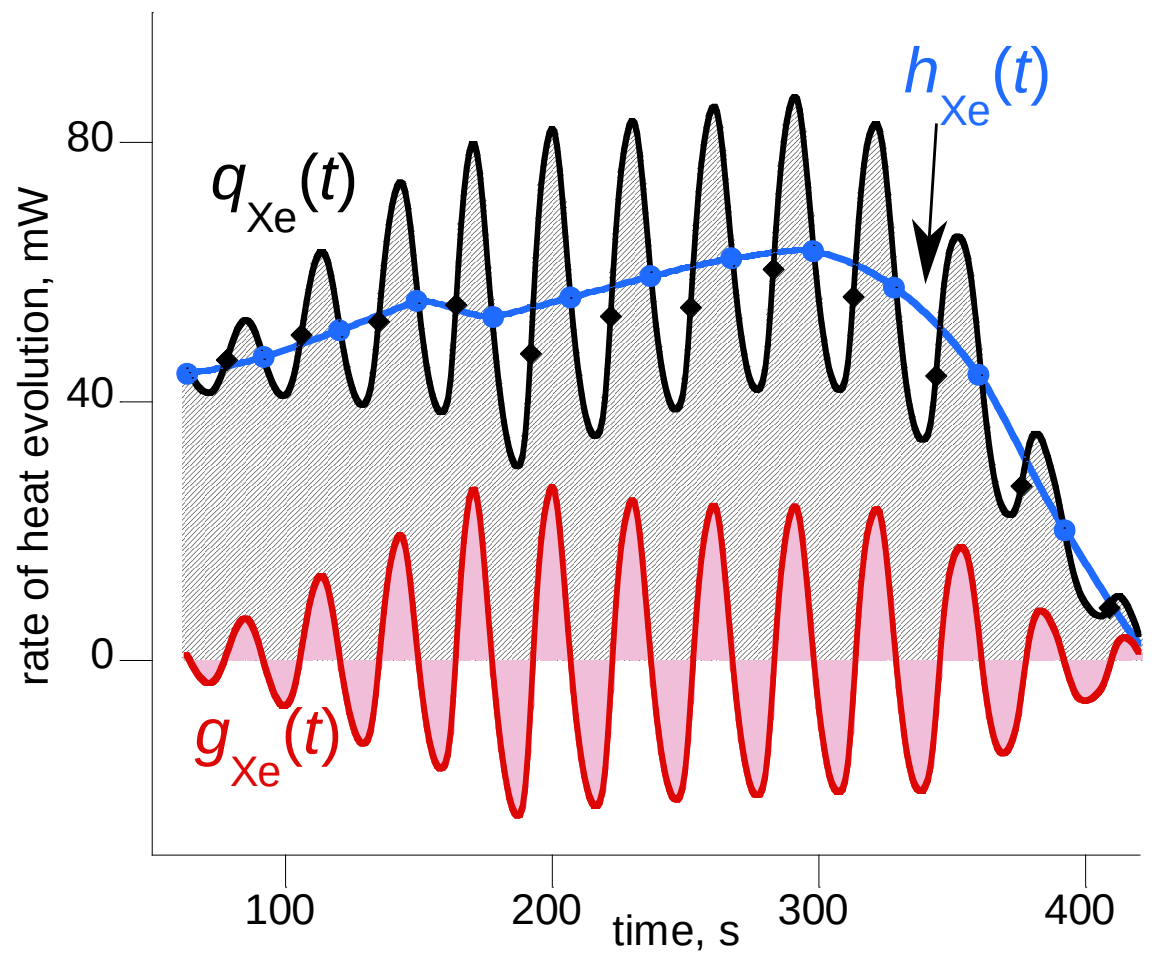

**Figure 6.** Extracting of the oscillatory component $g_{\mathrm{Xe}}(t)$ from non-equilibrium thermokinetic oscillations $q_{\mathrm{Xe}}(t)$ [from Dataset (2)] via a pointwise subtraction from it of a mean value curve $h_{\mathrm{Xe}}(t)$.

**4.2. Linear dependence of individual peaks' intensities on FIP.** Figure 7A represents the oscillatory components, $g_{\mathrm{Ne}}(t)$, $g_{\mathrm{Ar}}(t)$ and $g_{\mathrm{Xe}}(t)$, extracted from the three original time series shown in Figure 5 A for Ne, Ar and Xe. For each $g(t)$ curve, the total areas enclosed between the curves and the x-axis are necessarily zero. Nevertheless, the intensities of oscillations clearly increase on going from Ne to Xe, in the same manner as it is evident in the $q(t)$ time series of Figure 5 A. Unlike the original $q(t)$, however, the $g(t)$ curves oscillate around the zero, enabling an adequate resolution of exo and/or endo peaks to be made. Figure 7 B illustrates the principle of pairwise equivalence of the successive peaks using the Kr curve, $g_{\mathrm{Kr}}(t)$, as example. Accordingly, within individual periods, marked from 1 to 8, the sum of areas of the black and the cyan peak is zero. Still, the absolute values of those peaks change between periods from 1 to 8, as well as, crucially, across the $g(t)$ curves from Ne to Xe. With respect to the latter, the absolute values turn out to be a linear function of the inerts' FIP. Regarding the periods corresponding to each other in the $g(t)$ curves, Figure 7 C and D show the linear dependencies of the peak areas on FIP, plotted separately for individual periods from 1 to 8. The grey and cyan triangles represent exo and endo peak areas, respectively fit with the solid and the dotted straight lines. In most cases both the solid and dotted line are practically identical, in agreement with the pairwise equivalence of the exo and endo effects mentioned above.

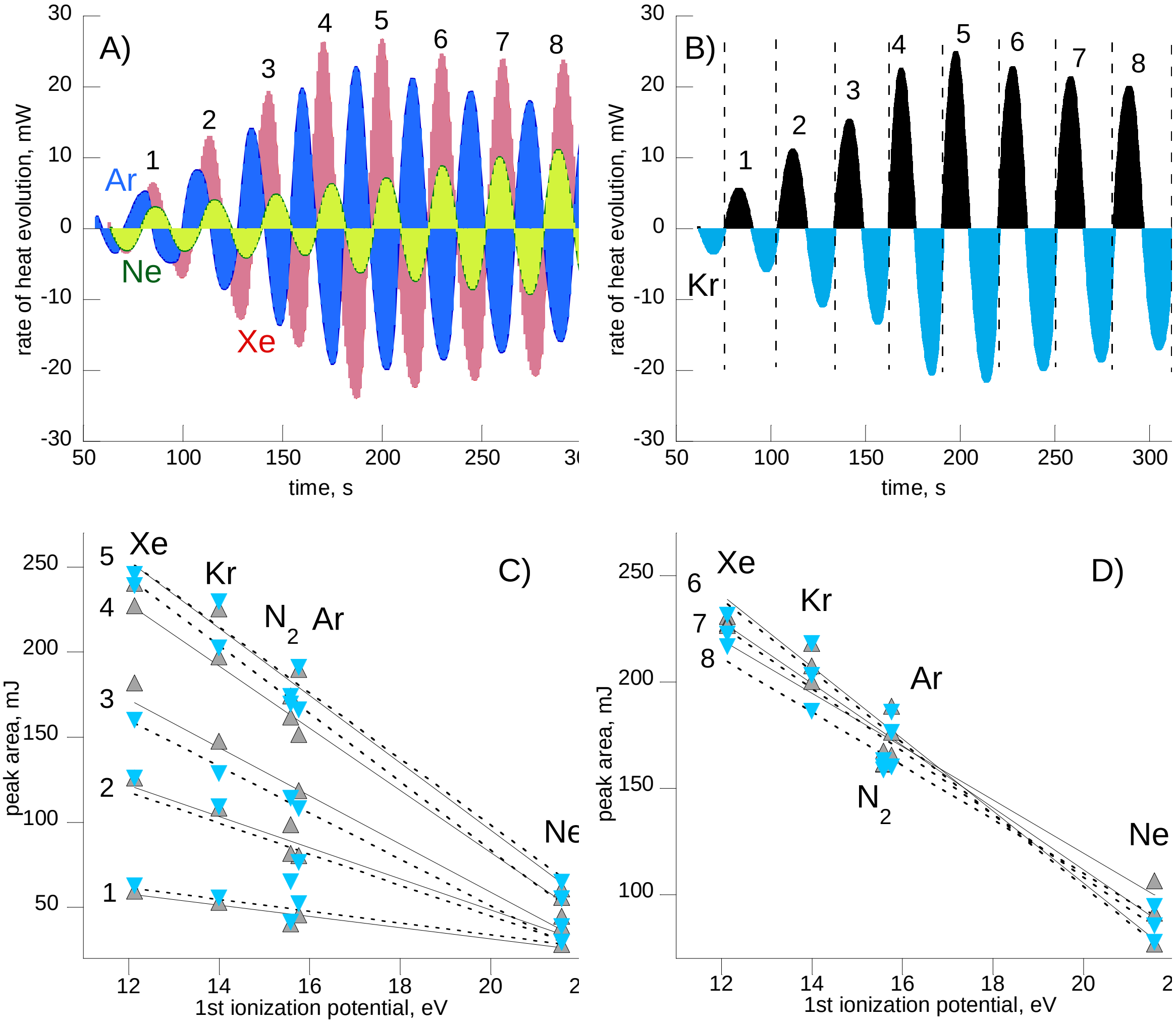

**Figure 7**. **A)** The extracted oscillatory components of the original thermokinetic time series of Fig. 1 A (for Ne, Ar and Xe). Arabic numerals mark successive periods. **B)** In each period (1 to 8) the area of exothermic (black) peak equals in absolute values to that of its endothermic successor (cyan) (data for Kr). **C)** and **D)** Linear dependence of the intensity of peaks in each period, from 1 to 8, on the first ionization potential (FIP) of the inerts used. The upward grey and the downward cyan triangles represent the absolute values of areas of the exo and endo peaks and their linear fit is shown with solid and dotted lines, respectively.

**4.3. Definition of the mean amplitude**. An explicit dependence of the amplitude on the FIP is now evident, but we seek a single amplitude-related descriptor for every $g(t)$ curve to be correlated with FIP. We begin by considering the integral with variable upper limit of the absolute value of the $g(t)$ curve:

$$G(t) = \int_0^t |g(t)| dt \tag{17}$$

(for simplicity, we refrain from using a "dummy variable" in the integrand). Figure 8 A shows the $G(t)$ curves plotted in juxtaposition to shaded areas under their original $|g(t)|$ functions. For clarity, the figure only shows the data for Ne and Xe,

$$G_{\mathrm{Ne}}(t) = \int_0^t |g_{\mathrm{Ne}}(t)| dt \,, \tag{17a}$$

$$G_{\mathrm{Xe}}(t) = \int_0^t |g_{\mathrm{Xe}}(t)| dt. \tag{17b}$$

Note, that a large segment in either $G(t)$ curves can be approximated by a straight line,

between $t1$ = 200 s and $t2$ = 350 s (cf. vertical lines in Fig. 8 A). Figure 8 B represents these linear approximations plotted for each of the inerts. The slope of those lines increases on going from Ne to Xe, which implicates a relation to the inerts' FIP. Before going to details of this relation, however, we need to find a mathematical sense of this slope of the $G(t)$ approximation with respect to the $g(t)$ function (cf. Fig. 8 C).

Mathematically, the slope of the line approximating the $G(t)$ curve within an interval ($t1$, $t2$) corresponds to the mean value of the absolute value of the $g(t)$ time series within the same limits. To see this, note, that by virtue of the mean value theorem for integrals [14], the mean value MV for $|g(t)|$ within the interval ($t1$, $t2$) can be expressed:

$$\mathrm{MV} = \frac{1}{(t2-t1)} \int_{t1}^{t2} |g(t)| dt \quad (18)$$

and considering relation (17) we have

$$\mathrm{MV} = \frac{1}{(t2-t1)} [G(t2) - G(t1)]. \quad (19)$$

Since the linear approximation of $G(t)$ may be expressed as $G(t) = at + b$, so we obtain:

$$\mathrm{MV} = \frac{1}{(t2-t1)} (at2 + b - at1 - b) = a. \quad (20)$$

This is illustrated in Figure 8 C, showing the fragment of the extracted oscillatory component $g_{\mathrm{Xe}}(t)$ (dashed red) and the linear approximation of the $G_{\mathrm{Xe}}(t)$ (in blue). The actual linear equation obtained for the latter is shown at the top of the figure and reveals its slope of 14. 864. On combining (18) and (20) we have:

$$a(t2 - t1) = \int_{t1}^{t2} |g(t)| dt \quad (21)$$

In Figure 8 C, the LHS of (21) is represented as area of the grey-shaded rectangle of height $a$ = 14.864 mW and width ($t2$ – $t1$) = 150 s, so its area amounts to 2229.6 [mJ]. The RHS of (21), in turn, can be found on numerical integration of the $|g_{\mathrm{Xe}}(t)|$ from $t1$ to $t2$, yielding 2214.7 [mJ], and corresponds to the absolute value of the red-dashed area in Figure 8 C. Clearly, the absolute values of the two areas marked in Figure 8 C are equal within experimental error. Since the slope $a$ equals MV, so it is hereafter referred to as the mean amplitude. Figure 8 D evidences the strictly linear dependence of the mean amplitudes on FIP for the whole database of the nine experiments; the data for this figure are listed in Table 4. Thus the mean amplitude can be considered as single amplitude-related descriptor for a $g(t)$ curve. We reiterate, at the same time, that each $g(t)$ curve uniquely represents its experimental original $q(t)$, having the same frequency and basic topology. Hence the mean amplitude can be considered as the same kind of descriptor also for the original thermokinetic time series $q(t)$. The mean amplitude so defined, therefore, is the desired, amplitude-related parameter, that can be used as a measure of intensity for chemical oscillations far from equilibrium.

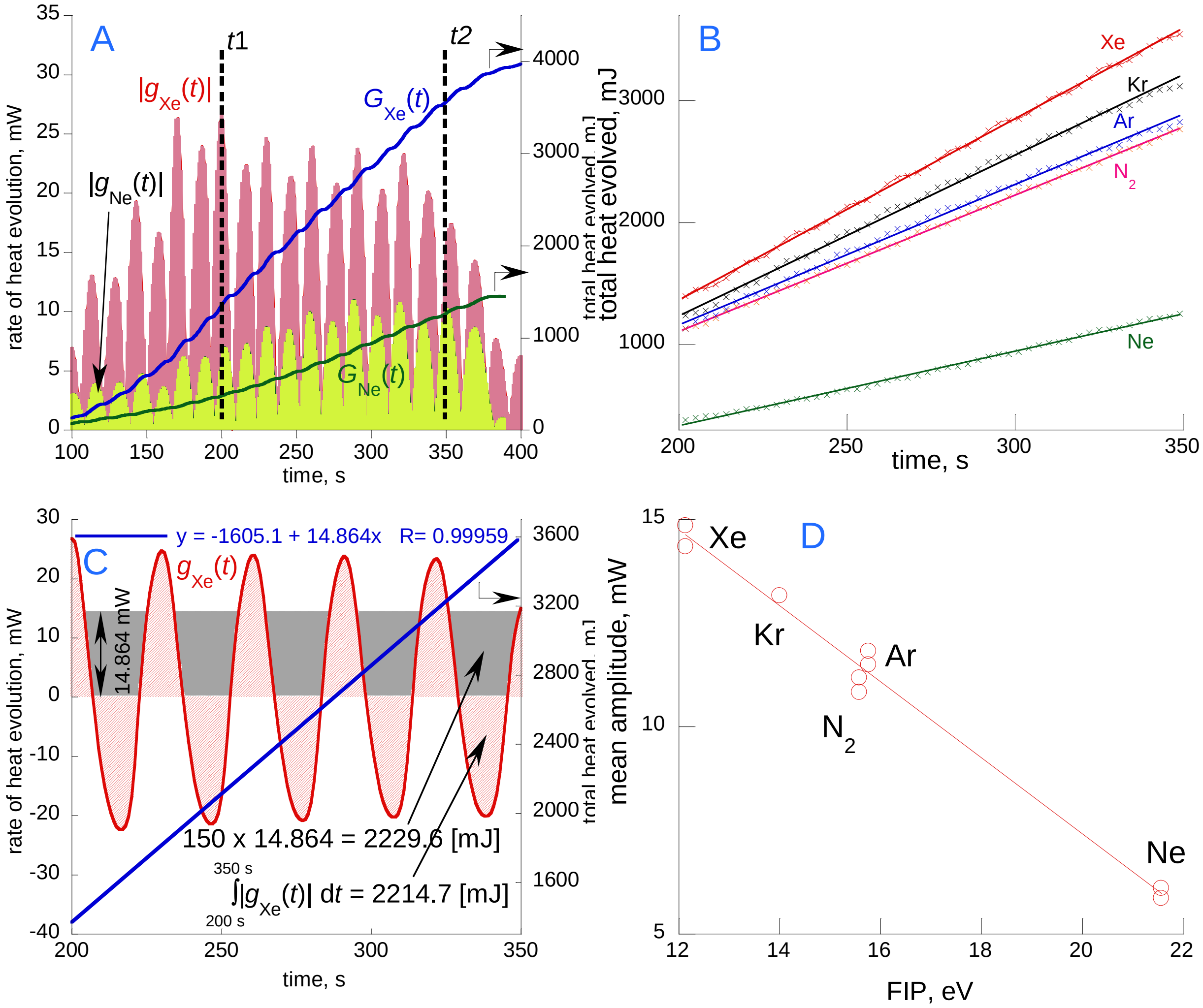

**Figure 8.** **A)** Shaded areas under the absolute values of the $g(t)$ curves for Ne and Xe, against the lines representing their cumulative integrals $G(t)$. **B)** The linear segments of the cumulative integrals $G(t)$ from $t1$ to $t2$ approximated by straight lines. **C)** Geometric interpretation of the mean amplitude as the height of a mean value rectangle (grey), equal to the slope of the linear approximation of $G(t)$ (14.864). **D)** Linear dependence of the mean amplitudes on FIP (RSD 0.3575, error bars not shown).

Table 4. The values of mean amplitude used for the plot in Figure 8 D. The absolute error is the ratio of the mean amplitude absolute residual (from least squares linear fit) and the average of the mean amplitude column.

| Inert gas admixture (Exp. No) | First ionization potential (FIP), eV | Mean amplitude, mW | Absolute residuals, mW | Absolute error, % |
|---|---|---|---|---|
| Ne(1) | 21.564 | 6.1101 | 0.12537 | 1.1318 |
| Ne(2) | 21.564 | 5.8700 | 0.11473 | 1.0358 |
| Ar(1) | 15.759 | 11.501 | 0.19760 | 1.7839 |
| Ar(2) | 15.759 | 11.827 | 0.52360 | 4.7270 |
| Kr | 13.999 | 13.164 | 0.24805 | 2.2393 |
| Xe(1) | 12.130 | 14.864 | 0.22963 | 2.0731 |
| Xe(2) | 12.130 | 14.342 | 0.28637 | 2.5853 |
| $N_2$(1) | 15.576 | 11.180 | 0.29107 | 2.6277 |
| $N_2$(2) | 15.576 | 10.839 | 0.63207 | 5.7063 |

## 5. Discussion.

**5.1. Co-adsorption of inert gases with hydrogen on Pd surface.** Considered together, the linear dependencies in Figs 5 B and 8 D both demonstrate a mathematically precise effect of the inert gases, Ne, Ar, Kr, Xe, or $N_2$, exerted when either is admixed to $H_2$ prior to its admission on Pd in the H/Pd sorption process. The intensity of the ensuing oscillations turns out to be inversely proportional to the inert gases' FIP, pointing out to their adsorption on Pd to be involved in generating the thermokinetic oscillations. The inert gases adsorption on metals is determined by the van der Waals forces, themselves a strong function of FIP [16]. As the FIP of noble gases decreases down the periodic table, so the strength of their adsorption on Pd increases in the same order [17]. Hence the stronger adsorption of the inert gas admixture, the more intensive the thermokinetic oscillations in the H/Pd sorption. Further supporting this view are our previous gravimetric results for adsorption of Ar on the Pd powder [18]. The Ar coverage on Pd turned out to be a linear function of its pressure within a range from zero to 700 Tr (cf. Figure 5 in Ref. [18]), with the straight line passing through the origin, and the adsorption being reversible in vacuum. Also, a parallel experiment reveled a significant co-adsorption of Ar with hydrogen on Pd [18]. Hence the strength of the Ar adsorption on Pd is independent of coverage, within a wide pressure range. Similar linear relations may be expected for other inert gases as well, especially for a lower coverage. Therefore, the relative rates of co-adsorption of the various inert gases with hydrogen on Pd should be solely a function of their strength of van der Waals attraction with Pd, determined by FIP. Since the initial partial pressure of the inert gases was set up the same in all experiments in Dataset (2), so the linear dependence on the FIP of the mean amplitude indicates that the intensity of those oscillations must be a function of the rate of co-adsorption of the inerts with hydrogen on Pd. In turn, the exact exo-endo balancing within each period (cf. Fig. 7) points out to a strong element of periodicity involved on the part of the coadsorption itself. Accordingly, the coadsorption of the inert gases on Pd should be viewed a succession of adsorption/desorption cycles, rather than as their steady coverage on the Pd surface. These adsorption/desorption cycles of the inerts may occur at the same frequency as the sorption process of hydrogen. This seems to be confirmed by preliminary results showing that the thermokinetic oscillations are indeed accompanied by fluctuations of total pressure in the gas phase of the same frequency [2]. More recent evidences of the correlation between the pressure and the thermokinetic oscillations are shown in Figure 12 which will be discussed in more detail in Section 5.4. To summarize this point, a mechanism of the inert gases' intervention, determining the intensity of oscillations in the sorption process, must presumably be periodic, likely involving adsorption/desorption cycles of the inerts on the Pd surface.

**5.2. Nonlinear hydrogen uptake.** From the dynamic point of view, our flow-through microcalorimetric system (the microcalorimetric cell, surrounded by heat sink, and the reaction carried out in flow-through mode) is a dissipative one. The reaction of hydrogen with palladium is accompanied by the evolution of heat that can be immediately dissipated to the heat sink in the instrument. The hydrogen reagent is provided at a constant flow rate to the Pd sample that resides in the microcalorimetric cell. Nevertheless, the system behaves like a microcalorimetric clock, and therefore, assuming that the rate of heat evolution measures the rate of uptake of hydrogen by Pd, an "escapement" mechanism must exist to make the hydrogen uptake periodic.

The periodicity of hydrogen uptake, as a part of an oscillatory mechanism inherent to the H/Pd sorption process, has been postulated in Ref. [4]. It has been based on combined measurements of the heat evolution and the electric conductivity of palladium powder, jointly recorded during the oscillatory sorption of deuterium in Pd. Repeatedly it was observed, that significant electric perturbations accompanied exclusively the ascending segments of the ensuing waveforms, but were evidently absent in the descending parts. Thus the electric instabilities coincided with the phases of increasing rate of heat evolution, manifestly disappearing during the phases of decreasing heat production. This is illustrated in Figure 9, that shows the electric perturbations, $\mathrm{d}I/\mathrm{d}t$, against the oscillatory component $g(t)$, as already extracted from the D/Pd thermokinetic oscillations (part of Dataset (1)). Note, that the coincidence with electric instabilities is more evident when the thermokinetic oscillations are fully developed. Accordingly, it was suggested in Ref. [4] that the periodic electric instabilities might be correlated with a periodic dissociation of molecular hydrogen on the Pd surface and immediate penetration of atomic hydrogen into the bulk of palladium.

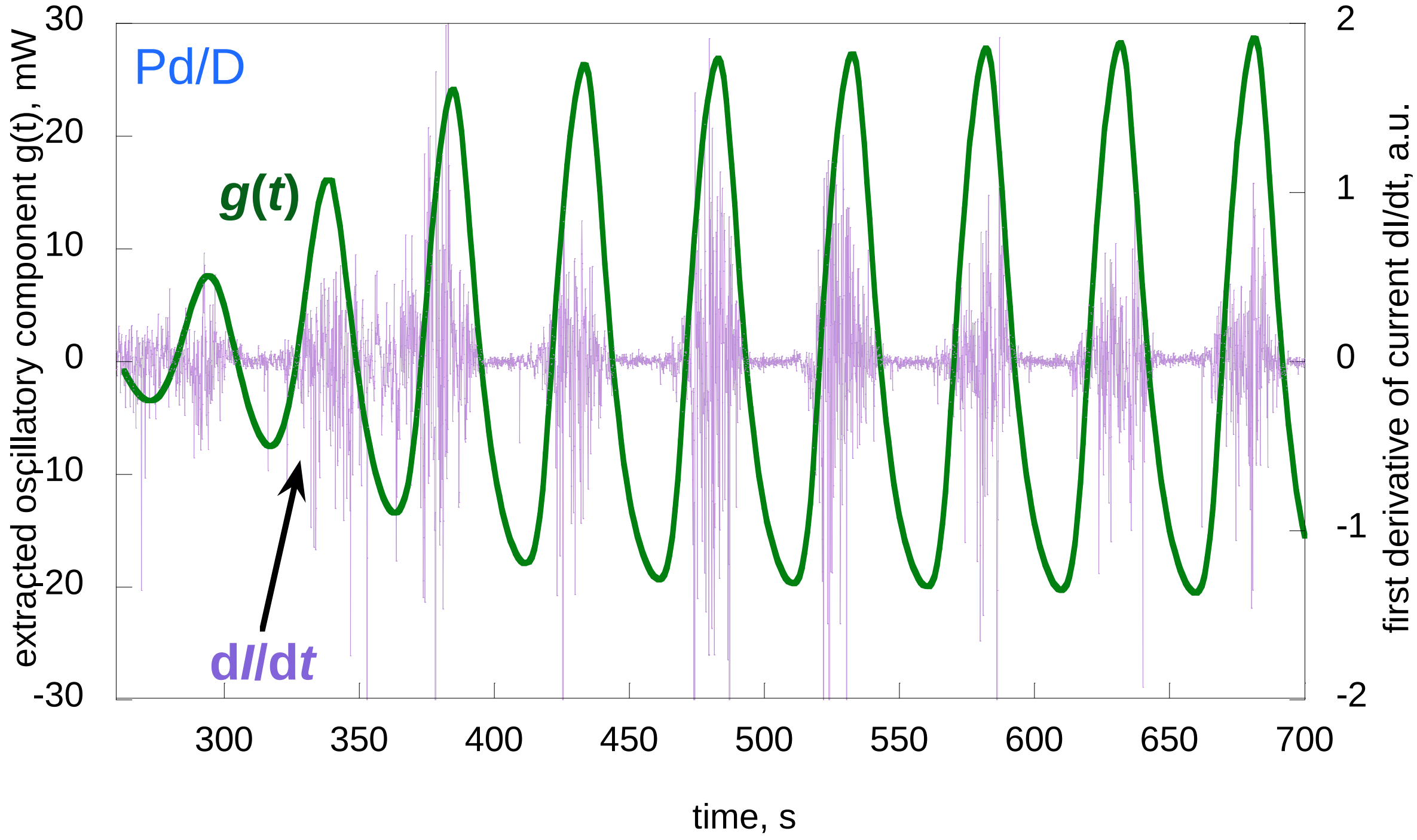

**Figure 9**. Correlation of electric instabilities with phases of increased heat evolution during oscillatory sorption in the D/Pd system. The green line represents the extracted oscillatory component of the experimental $q(t)$ time series.

The proposed mechanism [4] assumes a spontaneous formation of spatio-temporal heterogeneity of the Pd surface. Accordingly, the hydrogen molecules are initially adsorbed on the Pd surface forming a growing domains without immediately dissociating, that is, not until a certain critical amount of the $H_2$ ad-molecules is being amassed. Since this phase sees only a molecular adsorption, so its thermal effect will consistently be weak. Only on reaching a threshold coverage, the dissociation into atomic hydrogen species takes off and continues until the domain just formed is depleted. Crucially, this is a phase accompanied by an intensive heat evolution. At the same time, it is also seeing the dissociated hydrogen leaving the surface and penetrating into the bulk of Pd, thus concluding a full cycle. Hence the two phases make up a cycle which by repeating itself creates a kernel for the oscillatory rate of heat evolution. It can be viewed as a micro-oscillator confined to a minute compartment (μm-sized domain) formed spontaneously on the surface of Pd grains within powdered sample. The spontaneous formation of spatio-temporal patterns of domains, containing several thousand atoms, on the surface reactions has been reported by several researchers [19-28]. The whole reacting system may then be considered as a network of micro-oscillators, which can be observed microscopically [25-27].

**5.3. The self-sustainability of the oscillatory H/Pd sorption.** Figures Figs 5 B and 8 D show that, in the oscillatory sorption of hydrogen in Pd, both the frequency and the amplitude are determined by the kind of inert used as the admixture to the flow of hydrogen prior to its contact with Pd. On the other hand, they do not affect the change of enthalpy in the formation of the resulting palladium hydride $H_xPd$. The enthalpy change may be considered as an initial condition of the oscillatory sorption, since the H/Pd ratio, and thereby the enthalpy change, are strictly determined by the mass of Pd sample, as well as the temperature and pressure. Hence the initial conditions are not affected by the kind of inert admixtures. These aspects seem to satisfy the definition of oscillatory sorption as the self-sustained oscillations, in a sense, that the frequency and amplitude of oscillations are not determined by the parameters of energy source which in this process is the interaction of Pd + H. Using the microcalorimeter, it is possible to measure both the oscillatory parameters (frequency and amplitude) and the heat evolution accurately and with high reproducibility. Figure 10 demonstrates the reproducibility of the experimental recordings of thermokinetic oscillations in the H/Pd system using as an example two sorption cycles $q_1(t)$ and $q_2(t)$ carried out at the same conditions with $N_2$ as the inert admixture (comp. Fig. 1(f) in Ref. [2]). The figure also shows that in spite of markedly different initial periods both time series reach the same limit cycle which confirms the self-sustainability of the oscillations.

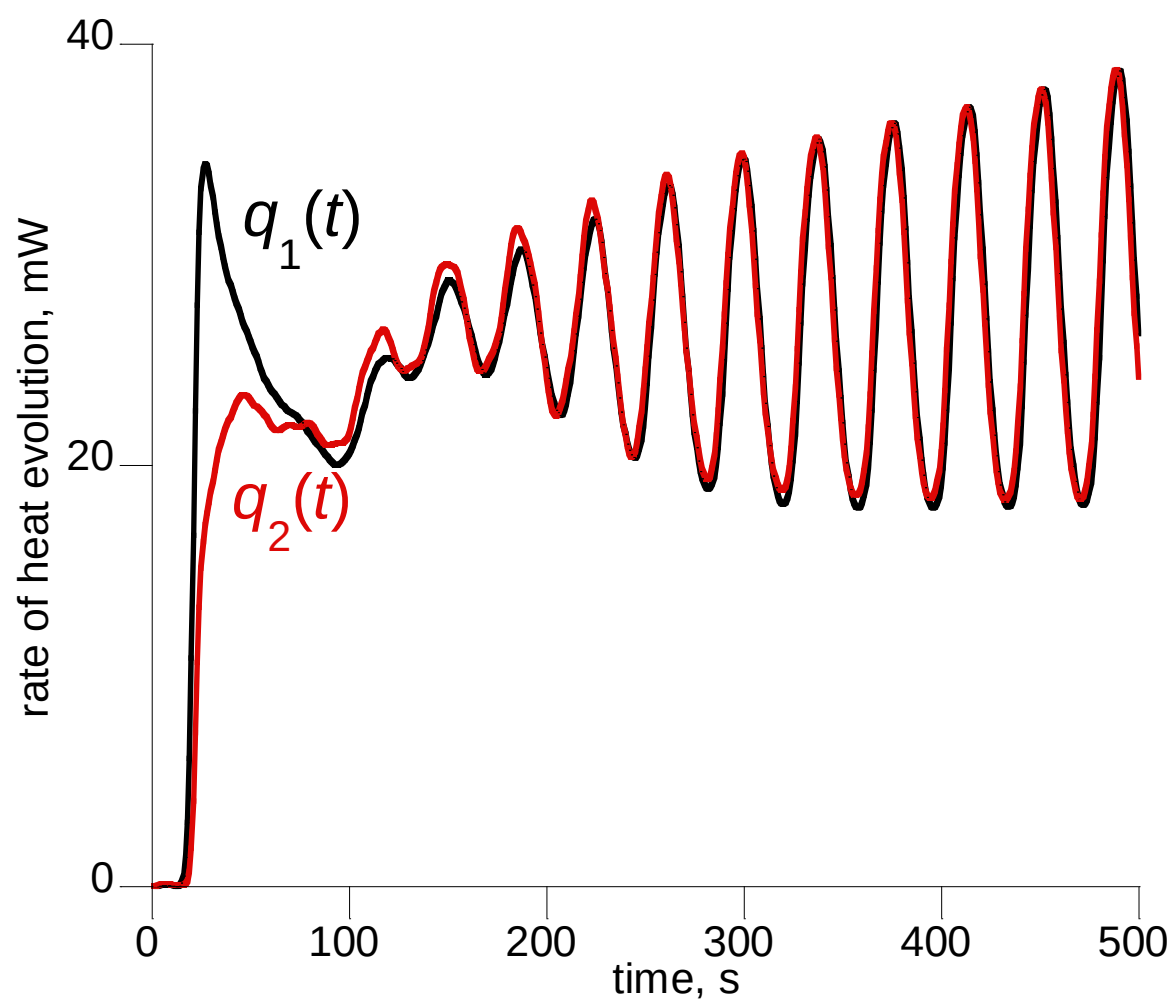

**Figure 10**. Reproducibility of the experimental recordings of thermokinetic oscillations in the H/Pd system. For clarity only the first 500 s of both time series are shown. In spite of diverse incubation periods, both oscillations develop the same frequency of 0.0262 Hz and the same amplitude. The areas under the whole curves $q_1(t)$ and $q_2(t)$ are respectively 41724 and 41447 mJ. The data used here are a part of results published in Ref. [2].

The exact exo-endo balancing over single periods has been a crucial feature consistently observed in the extracted oscillatory components for both Dataset (1) (cf. Fig. 4) and Dataset (2) (cf. Fig. 7). Mathematically, this is the result of relation (8). In terms of dissipative systems, this feature means that the energy pumped in is strictly compensated by the dissipated energy within each single oscillatory period [29]. Figure 11 A shows a fragment of oscillatory component (similar to that in Fig. 4 D) with the overimposed cumulative integral curve (in magenta), hitting the zero any time a period starts anew. This coinciding of zero points can be seen more clearly in Panel C. It leads to a closed limit cycle in the phase space, shown in Panel (B), with a single limit cycle loop represented in Panel (D). Thus the self-sustainable nature of the oscillatory H/Pd sorption appears to be rooted in each individual period, since each period is represented by a closed curve in the phase space, which is symptomatic for self-sustainability of oscillations [6,10]. We reiterate, that in calorimetric terms, a curve in the phase space shown in Figs 11 B and D represents certain rate of the heat evolution, which is $g(t)$, as a function of the corresponding total heat evolved, that is the area under $g(t)$.

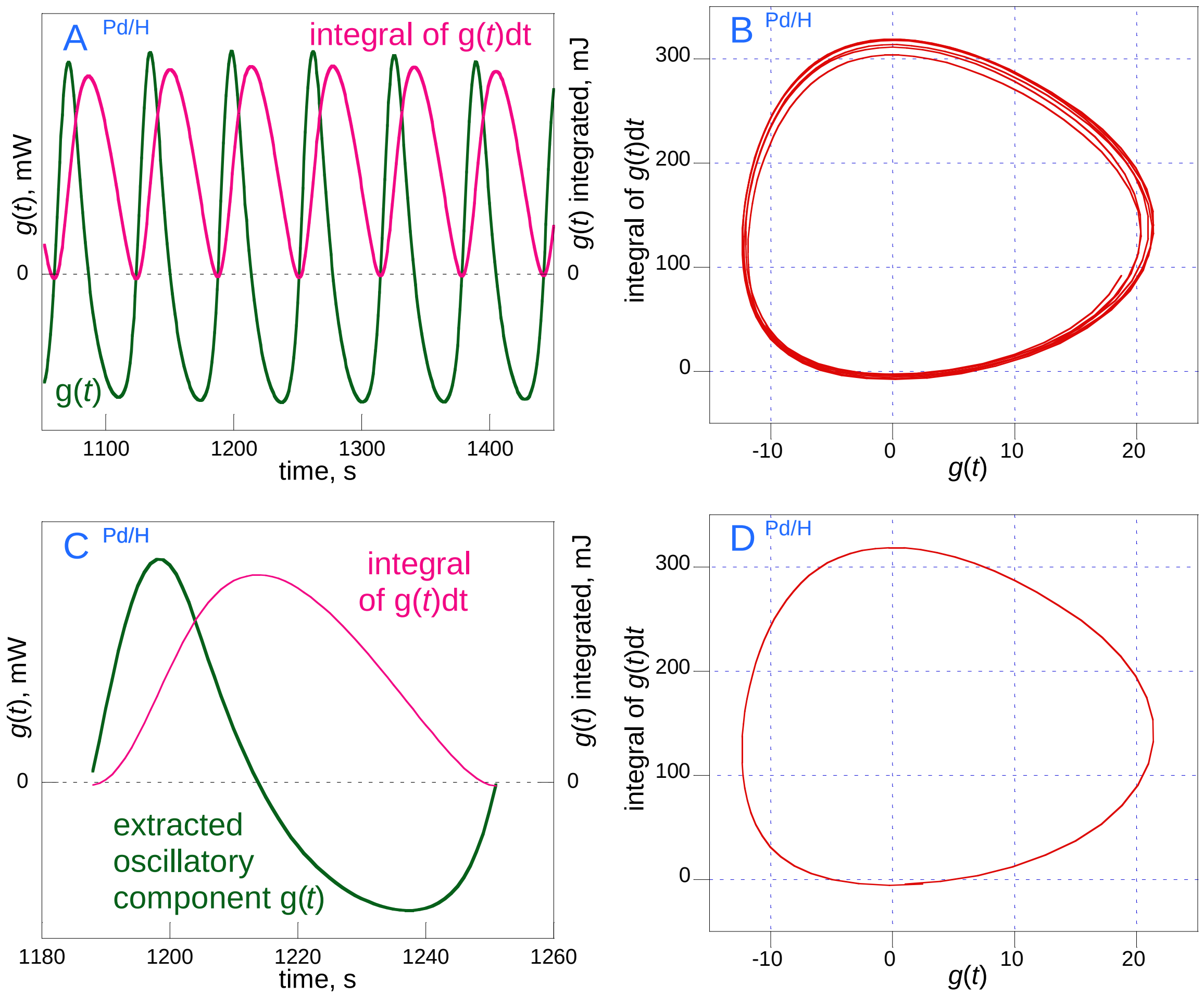

**Figure 11.** **A**) The extracted oscillatory component $g(t)$ (green) has its exo and endo peaks strictly compensating each other within each period, so that the integral curve of it (magenta) hits the zero at every period starting point. Resulting is a closed limit cycle shown in panel (**B**). Panels (**C**) and (**D**) respectively represent a single period and a single limit cycle loop in the phase space.

**5.4. The case for synchronization.** By this point we established the following three points: Firstly, the thermokinetic oscillations in the H/Pd sorption are self-sustained (cf. previous section). Secondly, in order for the H/Pd sorption process to act as a microcalorimetric clock, a nonlinear "escapement" mechanism of periodic hydrogen uptake from its continuous supply must be repeatedly operative in each individual period (cf. Section 5.2). Thirdly, the necessary co-adsorption of the inert gases with $H_2$ on Pd follows the same as the H/Pd sorption periodicity (cf. Section 5.1). The periodic uptake of hydrogen can be rationalized in terms the mechanism proposed in Ref. [4], but this mechanism does not clarify the nature of the inert gases intervention in the process. We will now attempt to reconcile this mechanism of periodic hydrogen uptake with the crucial role of the inert gases' co-adsorption in determining the oscillatory parameters. In doing so, we argue, that the role of the inert admixtures consists of providing a coupling medium for synchronization.

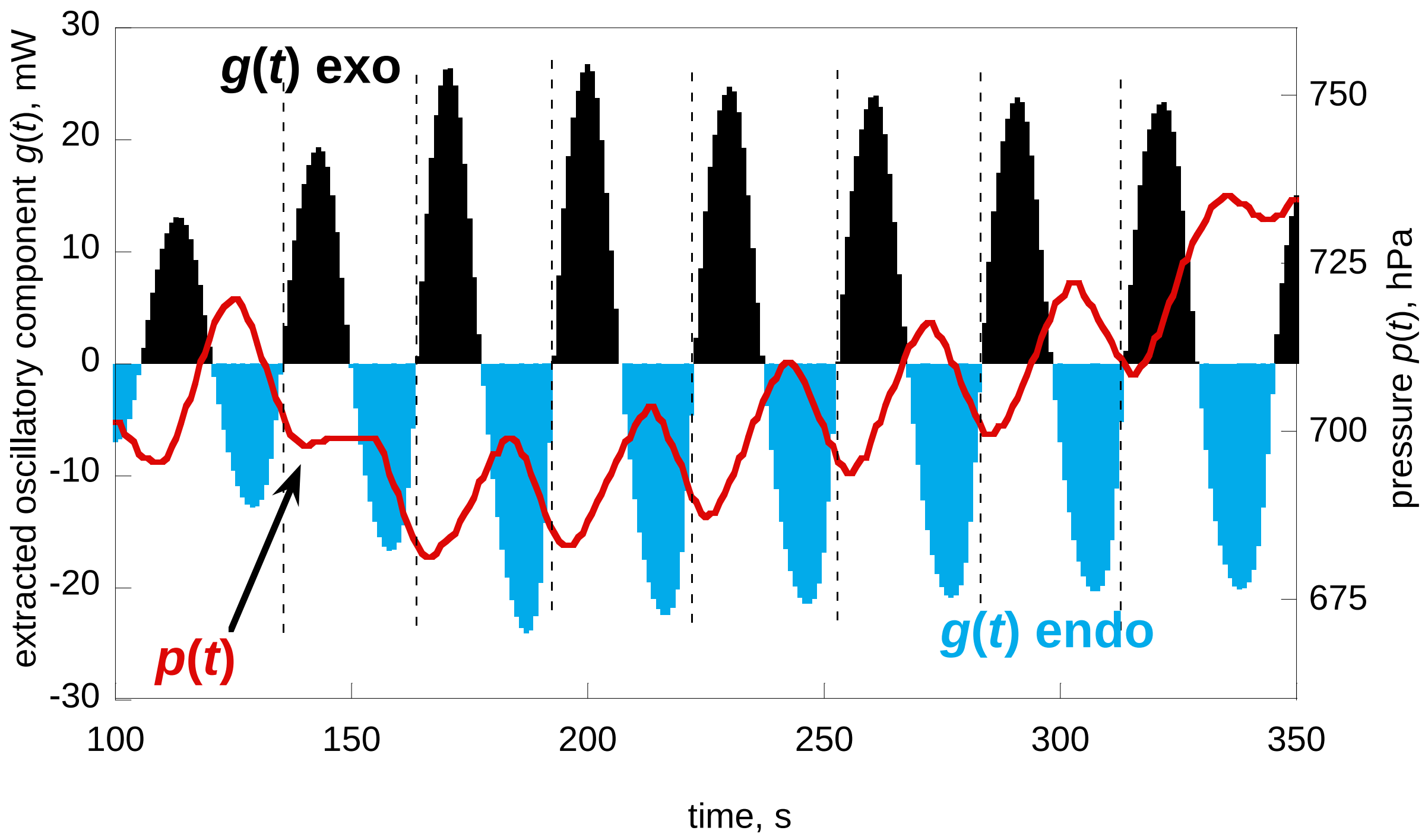

**Figure 12.** The fluctuation of pressure in the gas phase *p*(*t*) accompanying the oscillatory H/Pd sorption, with Xe as the inert gas admixture, represented by the extracted oscillatory component *g*(*t*).

According to the mechanism proposed in [4] the oscillatory H/Pd system can be viewed as an assemble of micro-oscillators (cf. Section 5.3). Unless synchronized, however, such micro-oscillations may not be observable in macro scale. The synchronization requires the micro-oscillators to be exposed to a weakly excitable medium (coupling medium) [6,30] In the powdered sample, the gas phase is the most likely to act as a coupling medium, relying on rapid transmission of pressure fluctuations for the weak excitations to be conveyed. This seems to be supported by our previous finding that the frequencies of pressure fluctuations in the gas phase during the H/Pd sorption are the same as those of the thermokinetic oscillations which they accompany [2]. It indicates, that the periodicity of adsorption/desorption cycles of inert (suggested by exo-endo balancing in Fig. 7 B), may help the gas phase to act as the coupling medium. Figure 12 shows the fluctuations of the gas phase pressure in one of the experiments from Dataset (2) using Xe as the inert admixture. The time series representing the pressure *p*(*t*) (in red) has been plotted against the extracted oscillatory component for this experiment. Clearly, the period of rising pressure coincide with the exothermic peaks (in black). The rising pressure may be related to the episodes of desorption of the inert from the Pd surface. The weakly coadsorbed inert gas on Pd surface may undergo desorption during the periods of the more intensive heat generation, thus causing an instantaneous increase of pressure in the gas phase. Similar results were also found for the other inerts used in Dataset (2). (A more detailed study of the relation between the changes of gas phase pressure and the thermokinetic oscillations in Dataset(2) will be published elsewhere.) Presumably, the strength of this effect of pressure rising due to the thermal induced desorption of the physisorbed inert gases will be dependent on their coverage on Pd. Naturally, it will be stronger for the heavier gases, like Xe, able to reach a higher coverage on Pd due to its lower FIP. Consequently, the coupling strength for synchronization may depend on the inert coverage of Pd surface which, we remember, is a linearly decreasing function of their FIP [16,17]. Hence the intensity of pressure fluctuation over Pd and the consequent coupling strength of synchronization should be increasing on going from Ne to Xe, in agreement with the relation presented in Figure 8 D.

A crucial feature of self-sustained oscillations is their ability to be synchronized [31]. It has been pointed out in Ref. [30] that a set of four conditions must be met in order for a system of self-sustained oscillators to be synchronizable. (1) The system should consists of a large number of self-sustained micro-oscillators, all repeating a relatively simple cycle. In this respect, we may consider the oscillatory heat evolution in the H/Pd sorption as a result of synchronization of a plethora of self-sustained micro-oscillators, all repeating a two-phase mechanism, in a sense as proposed in Ref. [4]. (2) Another condition is that those micro-oscillators must all undergo a weak interaction with a coupling medium, which in the H/Pd system may be conveyed by pressure fluctuations caused by periodic desorption of inert gases. (3) The third condition requires that a global coupling, rather than local, should exist between all the micro-oscillators involved. Since the pressure waves within a gas phase travel with the speed of sound, so all the micro-oscillators contained in the minute microcalorimetric cell (0.15 $cm^3$) can easily be responsive to those pressure fluctuations. (4) Finally, the micro-oscillators should be nearly identical. Taking for account that the reaction mechanism of the sorption in the H/Pd system must be the same in each and every micro-oscillating domain, so their frequencies may expectedly be sufficiently close to one another to warrant eventual synchronization. It seems therefore that the observed oscillatory rate of heat evolution in the H/Pd sorption satisfies the postulated nonlinear conditions for occurrence of the synchronization phenomena in the system.

## 6. Conclusions.

A mathematically rigorous method has been proposed to extract a pure oscillatory component from periodic time series recorded in oscillatory reactions that occur far from equilibrium. The concept was implemented using thermokinetic time series recorded microcalorimetrically in the oscillatory sorption of hydrogen and deuterium in Pd. Following discretization of the original data, the mean value theorem for integrals makes it possible to calculate a range of mean values, one for each individual period of thermokinetic time series. The obtained sequence is subsequently used as fit points to construct a flat curve modeling a hypothetical non-oscillatory heat evolution under otherwise the same reaction conditions. Crucially, the areas under both curves, i.e., the modeled flat and the experimental oscillatory one, are the same, attesting to the method's validity. This is due to the invariance of the molar heat of reaction, irrespective of whether the oscillations occur or not. The pointwise subtraction of this modeled base-line from the original curve yields a new time series representing the extracted oscillatory component. Since this new time series oscillates around the abscissa axis, so it is now possible to analyze the thermokinetic oscillations in terms of a succession of alternating exo and endo thermal effects. It reveals a pairwise exo/endo equivalence, i.e., each exo peak is followed by an endo peak that have the same area but of opposite sign. Hence, apart from the whole extracted oscillatory time series yielding zero on integration, also the exo and endo effects equilibrate one another within each period individually. This detailed exo/endo balance is a reflection of the mathematical relation expressed in equation (8). Physically, it seems to reflect an escapement mechanism operative at a very elementary level of the μm-size domains, viewed as micro-oscillators of which formation and synchronization enable the oscillatory kinetics of the sorption process to be calorimetrically detectable in macro scale.

It should be stressed, that the oscillatory components $g(t)$, extracted from the thermokinetic time series $q(t)$, retains both the frequency and the topology of the original to the extent that once extracted it does not cease to represent the characteristics of the experimental oscillations recorded. Effectively, the $g(t)$ time series obtained for Dataset (2) could reliably be subjected to further mathematical operations, making it possible for the mean amplitudes of the extracted components to be defined. In turn, the empirical finding of their linear dependence on the inert gas FIPs in Dataset 2 (cf. Fig. 8 D) confirms that the mean amplitude so defined can be used as a descriptor that unambiguously represents the intensity of thermokinetic oscillations as observed experimentally. In more general terms, however, the eventual successful correlation of the mathematical constructs with the real life reaction conditions may be considered as an experimental verification of both the concept and the mathematical procedure behind the proposed method of oscillatory component extraction. This method appears therefore to be confirmed as a mathematically rigorous treatment of oscillatory data.

## Acknowledgements

The authors acknowledge the financial support of the National Centre of Science (National Science Centre, Poland, grant no. 2012/07/B/ST4/00518) as well as the support and assistance of Microscal Energy Technology Ltd (Thatcham, UK).